\documentclass[10pt]{iopart}

\usepackage[backend=biber, style=phys]{biblatex}  
\expandafter\let\csname equation*\endcsname\relax
\expandafter\let\csname endequation*\endcsname\relax
\usepackage{amsmath, amsbsy}
\usepackage{graphicx}
\usepackage[unicode=true, bookmarks=true, bookmarksnumbered=true, bookmarksopen=true, bookmarksopenlevel=1, breaklinks=false,pdfborder={0 0 0}, backref=false, colorlinks=true]{hyperref}
\hypersetup{citecolor=blue,linkcolor=blue}
\usepackage{breakurl}

\usepackage[dvipsnames]{xcolor} 
\usepackage{lipsum,multicol} 

\usepackage [english]{babel}
\usepackage [autostyle, english = american]{csquotes}
\MakeOuterQuote{"}

\usepackage{subcaption}

\usepackage{graphicx}

\DeclareUnicodeCharacter{2009}{\,}

\begin{document}

\title{SPARC Tokamak Error Field Expectations and Physics-Based Correction Coil Design}

\author{N.C.~Logan$^1$,
C.E.~Myers$^2$,
R.~Sweeney$^2$,
C.~Paz-Soldan$^1$,
M.~Pharr$^1$,
N.~Leuthold$^1$,
M.~Nickerson$^2$,
J.~Halpern$^1$,
I.~Stewart$^1$,
}
\address{
$^1$ Columbia University, New York, New York 10027, USA

$^2$ Commonwealth Fusion Systems, Devens, MA 01434, USA

 }
 \ead{nikolas.logan@columbia.edu}
 
\vspace{10pt}
\begin{indented}
\item[]\today
\end{indented}

\begin{abstract}

Non-axisymmetric magnetic field coils have been designed to provide efficient error field correction and suppress edge localized modes in SPARC — a compact high-field tokamak that is presently under construction at Commonwealth Fusion Systems. These designs utilize the Generalized Perturbed Equilibrium Code's (GPEC's) representation of the multi-modal, non-axisymmetric plasma response to optimize the geometric coupling between 3D coil arrays and the desired core or edge plasma response. Error field correction coils are designed to couple to the plasma-amplified kink that dominates the drive of core resonances. The maximum allowable error field is projected to SPARC using an empirical scaling that is consistent with linear and nonlinear MHD modeling expectations. Asymmetric construction and assembly tolerances are then balanced against the corresponding kA-turns needed for correction to levels below the allowable limit. These physics-driven coil designs provide confidence in our ability to operate SPARC in new high field tokamak regimes without error field induced locked modes.

\end{abstract}

%
%
%
\maketitle
%
\ioptwocol


\section{\label{sec:Motivation} Motivation}

Non-axisymmetric error fields (EFs) in otherwise toroidally axisymmetric tokamak plasmas can cause macroscopic bifurcations from a plasma equilibrium confined on concentric flux surfaces to one with magnetic islands inside it, leading to a disruptive loss of plasma confinement. The risk of this outcome is manageable by controlling the magnitude of the EF source and by proactively applying known 3D fields to compensate for the intrinsic sources of asymmetry. The intrinsic sources of asymmetry can be asymmetries in the primary coils themselves, such as windings of the conducting tape or local lead junctions, or from installation (such as shifts and tilts of the coils). Compensating fields can be applied by toroidal coil arrays, as is now common practice in tokamak construction \cite{Robinson1995DevelopmentDIII-D, Buttery1999, Rott2009Electro-magneticUpgrade, Menard2010ProgressPlasmas, Kim2009DesignCoils, Wang2016, Foussat2011FromCoils}. The SPARC high-field tokamak being built by Commonwealth Fusion Systems (CFS) will utilize EF correction (EFC) to reduce a known EF, but it remains necessary to control EF magnitudes with strict tolerances for possible sources. This paper reports the control of EF risks through tolerances and EF risk mitigation strategies for SPARC tokamak operation. 


The projected thresholds for core EF locking are used here to detail the physics design considerations of the SPARC error field correction coils (EFCCs).  These coils were designed utilizing the principles of physics-led design \cite{Logan2020PhysicsTokamaks} within the highly constrained system of a compact tokamak. The principal plasma-physics aspects of the coil design capabilities are detailed here, with detailed engineering considerations being outside the scope of this report.

\section{\label{sec:SPARC} SPARC Plasmas}

This section briefly reviews the SPARC primary-reference-discharge-like (PRD-like) equilibria used to assess EF physics in this work. The SPARC PRD design point is described in much greater detail in Ref. \cite{Creely2020OverviewTokamak}.

For the purposes of this work, we will limit ourselves to two representative equilibria during a single SPARC discharge evolution. These are representative times from the L-mode (plasma current ramp-up) and H-mode (current flattop) phases of the discharge, taken from 7 and 14 seconds into the planned discharge evolution respectively.

\begin{figure}[h]
\centering{}\includegraphics[width=1.0\linewidth]{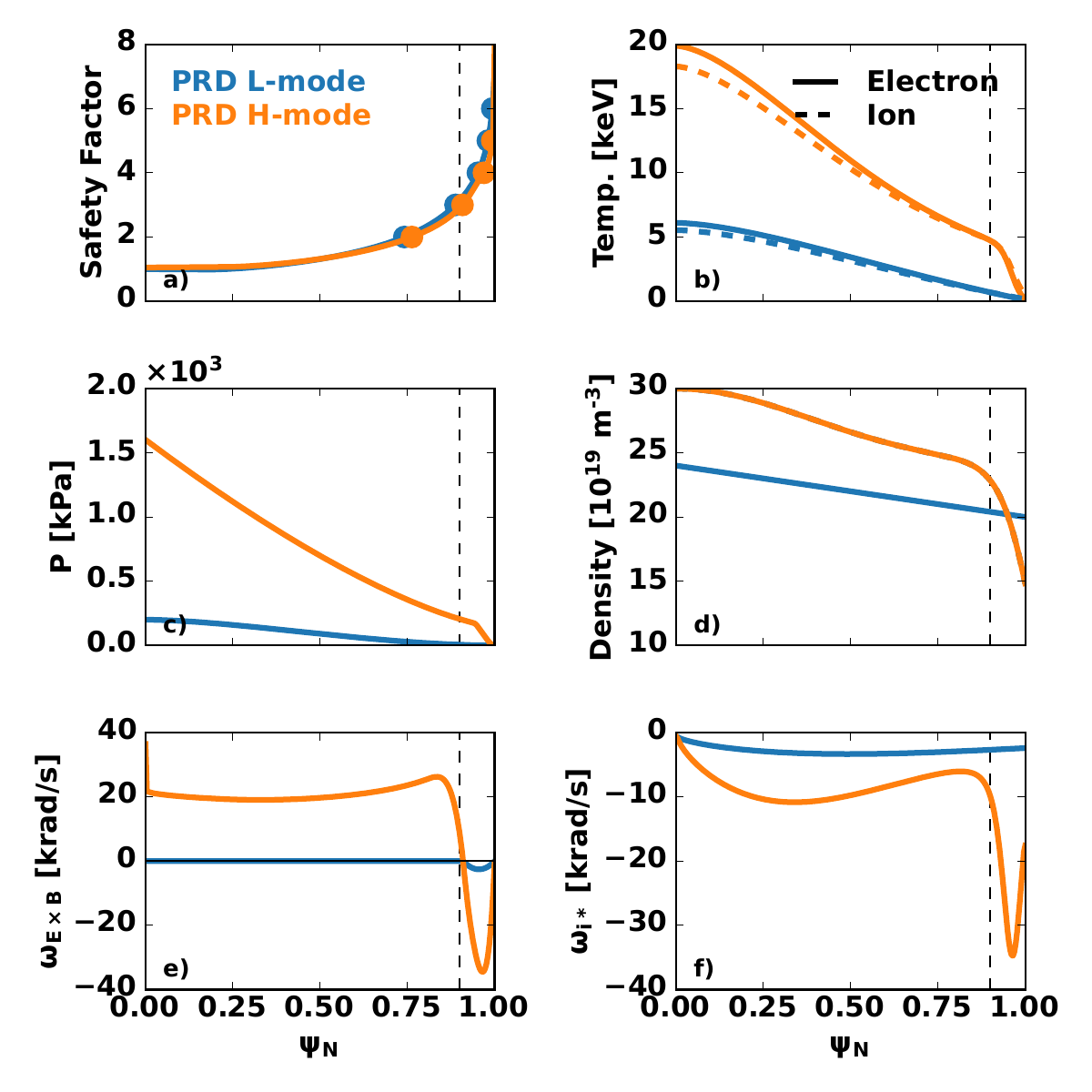}\protect\caption{SPARC PRD safety factor (a), temperature (b), pressure (c), density (d), $\mathbf{E}$$\times$$\mathbf{B}$ precession frequency (e), and ion diamagnetic frequency (f) profiles for the representative L (blue) and H (orange) mode  times of 7 and 14 seconds respectively. \label{fig:equil_summaries}}
\end{figure}

\begin{figure}[h]
\centering{}\includegraphics[width=1.0\linewidth]{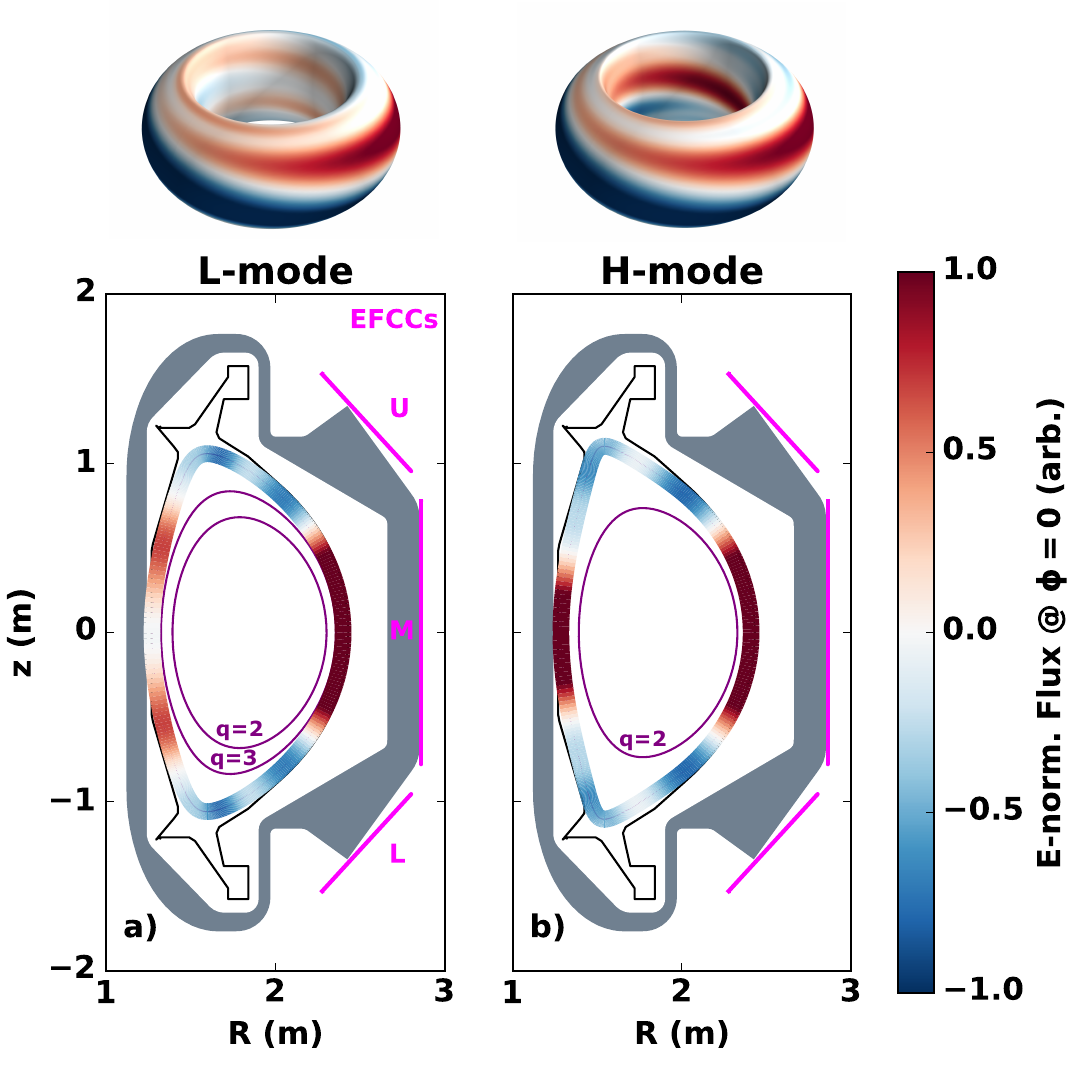}\protect\caption{Dominant mode structure of the representative SPARC PRD (a) L-mode and (b) H-mode times. The upper (U), midplane (M) and lower (L) EFCCs are shown in magenta. Core rational surfaces are shown in purple. \label{fig:dominant_modes}}
\end{figure}

The profiles from each time are shown in \fref{fig:equil_summaries} and shapes are shown in \fref{fig:dominant_modes}. 
In this modeled evolution, known as the CFS 2000-series, the MEQ suite's FBT code \cite{Hofmann1988FBTPlasmas} was used to make time-independent equilibria (i.e. the solutions do not include the induced wall currents) for a PRD-like scenario. Wave heating starts here at 8 s with the plasma current $I_p$ = 8 MA, and the double-null boundary shape, P' and FF' profile shapes are a close match to the PRD developed with TSC \cite{Creely2020OverviewTokamak}. 
The strike-point locations are different from the PRD due to the changes to the divertor topology that occurred after the PRD was formulated, but these are not expected to impact this EF work.
Importantly, self-consistent coil currents are available for all the toroidal field (TF), poloidal field (PF), central solenoid (CS) and divertor (DIV) coils in this 2000 series of equilibria.
Dipole constraints are placed on different sets of shaping coils to avoid inefficient coil current configurations where fields from adjacent coils cancel each other. 
These currents will come into play when assessing the EFs that arise from tilts and shifts of each coil.

The equilibrium, current and pressure profiles are all that is needed for the perturbative ideal MHD calculations that are the heart of this study. Inclusion of the neoclassical toroidal viscosity (NTV) in Sec \ref{sec:nonresonant} requires density, temperature, and $\mathbf{E}$$\times$$\mathbf{B}$ rotation profiles \cite{Logan2013NeoclassicalGeometry}. The profiles used here are shown in \fref{fig:equil_summaries}. They assume a pedestal width of 0.03 in normalized poloidal flux, $\psi_N$, a flat effective ion charge profile $Z_{eff}=1$, and a separatrix density of $1.5 \times 10^{20} \mathrm{m}^{-3}$ ($\sim15$\% of the Greenwald density \cite{Greenwald2002}).

The predicted $\mathbf{E}$$\times$$\mathbf{B}$ rotations for SPARC are shown in \fref{fig:equil_summaries}(e) with different assumptions for the L-mode and H-mode plasmas. For H-mode, the core region is assumed to be dominated by the toroidal rotation as predicted by the multi-machine empirical scaling developed in \cite{Rice2007InterTokamaks}. 
The toroidal scaling gives the Alfv\'en Mach number $M_\mathrm{A}$ at the $q=2$ surface as a function of toroidal beta $\beta_\mathrm{t}$ and the cylindrical safety factor $q^*$ as $M_\mathrm{A} = 0.65 \cdot \beta_\mathrm{T}^{1.4} \cdot q^{*2.3}$, which is translated into a toroidal rotation.
While this scaling was developed in purely wave-heated H-mode plasmas without torque input at the $q=2$ surface, we assume that it is a fair approximation for the intrinsic toroidal rotation in the whole core region from $\psi_N = 0$ to $\psi_N(q=2) + 0.05$. 
Since no scalings laws for intrinsic rotation exist in L-mode, we assume $\omega_\mathrm{ExB} = 0$ in the core.
The edge toroidal rotation for both, L-mode and H-mode, is set by two constraints: (1) that the radial electric field $E_\mathrm{r}$ well in the pedestal region is dominated by the main ion diamagnetic rotation (see $\omega_\mathrm{i*}$ in \fref{fig:equil_summaries}(f)) \cite{McDermott2009EdgePlasmas, Viezzer2014ParameterPlasmas} and (2) that there is no radial electric field at the separatrix $E_\mathrm{r}(\psi_N=1) = 0$ \cite{Bai2022ToroidalStability, Plank2023ExperimentalUpgrade}.
Both are approximate conditions set based on empirical experience rather than strict physical requirements, and more work on rotation prediction would further improve this model.
The corresponding intrinsic torque for SPARC, also predicted from scalings in \cite{Rice2021DimensionlessPlasmas}, is $T_0 = 4$ N$\cdot$m. 

\begin{table}[t]
  \centering
  \begin{tabular}{|l|c|c|c|c|}
    \hline
    Param & Units & L-mode & H-mode & ITER \\
    \hline
$t$   & s &  7.0  &  14.0 & - \\
$B_{T}$   & T &  12.16  &  12.16 & 5.3 \\
$R_{0}$   & m &  1.85  &  1.85 & 6.2 \\
$n_e$   & $10^{19}$ m$^{-3}$ &  17.3  &  28.8 & 9.8 \\
$\beta_N$   & - &  0.17  &  0.98 & 1.8 \\ 
$\ell_i$   & - &  0.74  &  0.72 & 1.0 \\  
$q_{95}$   & - &  3.95  &  3.69 & 3.1 \\
$q_{90}$   & - &  3.14  &  2.90 & 2.7 \\
    \hline
  \end{tabular}
  \caption{\label{tab:equil_scalars}Scalar quantities for representative SPARC equilibria and the ITER Baseline Scenario (H-mode) for context.}
\end{table}

Important scalar quantities for each equilibrium are given in Table \ref{tab:equil_scalars}, alongside the ITER Baseline Scenario values used in Ref. \cite{Logan2020RobustnessScaling} scalings for context. Notably, the SPARC plasmas are higher field and density while lower in size and $\beta_N$ than ITER.  

\section{\label{sec:EFs} Error Field Requirements for SPARC}

This section details how error fields are quantified in order to asses the risks they pose for inducing core locked modes. Primary importance is given to the most dangerous spectrum of the error field, greatly reducing the complexity of the problem. The risk associated with a given amount of this dangerous error field is taken from empirical scalings that use the same metric. The limitations of this single-mode metric approach are then discussed, considering the secondary effects of the less resonant error field components. A model for the expected error fields in SPARC is presented, allowing for detailed risk assessments given coil designs and there associated assembly tolerances.

\subsection{\label{sec:DominantMode} The Dangerous Component of Error Fields}

Error fields in this work are quantified by the core dominant mode overlap, normalized by the toroidal field on axis. This metric has been used to develop multi-machine scaling laws for the projection of error field thresholds to ITER, as will be discussed in the next section. This metric is detailed extensively in Refs. \cite{Park2008ErrorITER,Park2011ErrorHandedness, Paz-Soldan2014TheImportanceDIII-D, Logan2016IdentificationReluctance, Logan2020EmpiricalTokamaks, Logan2021PhysicsTokamaks}, and reviewed here succinctly for completeness in order to discuss SPARC-specific features of the metric.

To quantify the importance of any given non-axisymmetric field spectrum, one must first identify a physics metric for the impact the field has in the plasma. 
Historically, the error field community has been focused on the potential of such fields to drive core tearing modes. 
Linear tearing theory provides thresholds in the resonant radial field at rational surfaces $q=m/n$ \cite{Cole2006, Cole2008, Fitzpatrick2012NonlinearPlasmas, Fitzpatrick2022InfluencePlasmas}, where $q$ is the safety factor, $m$ is the integer poloidal mode number, and $n$ is the integer toroidal mode number. 
The resonant field is fully shielded by the ideal MHD plasma response prior to EF penetration, and the threshold is thus commonly characterized by an external tearing drive, $\Delta'_{x}$, \cite{Fitzpatrick1991, Fitzpatrick1993, Fitzpatrick2012NonlinearPlasmas} proportional to the delta-function shielding current on a rational surface. This work uses the effective resonant field metric, defined as the amount of resonant field shielded by the resonant sheet current, which is proportional to $\Delta'_{x}$ on a given surface \cite{Boozer2006PerturbedEquilibria, Park2007ControlTokamaks}.   

Within the ideal MHD framework of GPEC \cite{Park2007ComputationEquilibria, Park2009ImportanceTokamaks, Park2017Self-consistentTokamaks}, the effective resonant field on each rational surface in the computation domain is linearly related to the external energy-normalized flux such that,
\begin{equation}
\tilde{\mathbf{\Phi}}_{r}=\boldsymbol{{\cal C}}\cdot\boldsymbol{\tilde{\Phi}}_{x}.\label{eq:C matrix equation}
\end{equation}
Here $\tilde{\mathbf{\Phi}}_{r}$ is an $R\times1$ matrix vector of the energy normalized effective resonant field at each rational surface and $R$ is the number of rational surfaces in the computational domain within the plasma region of interest.
The vector $\boldsymbol{\tilde{\Phi}}_{x}$ is an energy normalized $M\times1$ external flux vector of poloidal Fourier harmonics $\{ \tilde{\Phi}_m, \: m_{min} <m<m_{max}\}$ \cite{Park2011ErrorHandedness, Logan2016IdentificationReluctance}, $M$ is the number of harmonics included in the spectra computation, and $\boldsymbol{{\cal C}}$ is a $R\times M$ coupling matrix.

For EFC threshold scaling purposes, where the main concern is core island penetration, a truncated coupling matrix is used where only rows corresponding to rational surfaces in the plasma that are contained within 90\% of the poloidal flux ($\psi_N<0.9$) are kept.
Decomposing this positive definite matrix using singular value decomposition gives positive singular values ranking the right singular vectors (RSVs) by how much resonant flux is induced by a unit vector of flux aligned with the RSV spectrum.
The first of these usually has a singular value much larger than the others and dominates the total resonant drive in tokamak scenarios of interest for fusion energy \cite{Park2008ErrorITER, Logan2021PhysicsTokamaks,Park2008ErrorITER,Park2011ErrorHandedness, Paz-Soldan2014TheImportanceDIII-D,Paz-Soldan2014TheSpectralDIII-D, Paz-Soldan2015DecoupledFields}, and we will thus refer to it as the dominant mode. The dominant mode identifies the most dangerous EF spectrum (the spectrum that drives the most effective resonant field per unit of EF on the plasma surface) and how much of the EF is aligned with this dangerous spectrum is referred to as the dominant mode overlap. 

The dominant modes for the representative L- and H- mode SPARC equilibria are shown in \fref{fig:dominant_modes}.  
Note that the dominant mode contains information about the helicity and shaping dependencies of the plasma response. 
The safety factor ($q$) profile and pressure profiles for the representative SPARC equilibria are shown in \fref{fig:equil_summaries}. 
The largest impact on the dominant mode structure changes between these two comes from the fact that the $q=3$ surface moves slightly outside the $\psi_N<0.9$ definition of core modes. 
As a consequence, the H-mode dominant mode is identically the $m/n=2/1$ coupling (note, $q=1$ surfaces have been excluded from the analysis, as their impact is assumed ignorable due to benign sawtooth redistribution of the core current).
The plasma pressure also plays a role in general, with larger pressures creating a higher sensitivity to low-field side EFs due to amplification from the ballooning kink response \cite{Logan2021PhysicsTokamaks}.
The SPARC H-mode $\beta_N$ is low enough, however, that the pressure does not have a large impact on the core dominant mode structure as compared to the L-mode.

Note that pressure impacts the singular value associated with the dominant mode (i.e. the magnitude of response) \cite{Lanctot2011, Wang2015}, but we will use a spectral metric to quantify the dangerous EF aligned with the dominant mode without the singular value. 
The core coupling dominant mode shape is a more robust feature of experimental equilibria than the singular values \cite{Park2008ErrorITER, Logan2021PhysicsTokamaks}, and a metric quantifying the amount of EF driving this mode has been chosen to be as robust to experimental uncertainty as possible. 
Minimizing this dominant mode overlap metric has been used to optimize EFC in active machines \cite{Park2011ErrorHandedness,Park2012SensitivityPlasmas,Paz-Soldan2014TheImportanceDIII-D,Paz-Soldan2014TheSpectralDIII-D, Yang2024TailoringTransport}. 
Calling the first the core resonant dominant mode $\hat{\tilde{\boldsymbol{\Phi}}}_{c1}$, the normalized overlap is,
\begin{equation}\label{eq:delta_definition}
    \delta = \frac{\left| \tilde{\boldsymbol{\Phi}}_{x} \cdot \hat{\tilde{\boldsymbol{\Phi}}}_{c1}\right|}{B_T}.
\end{equation}
Here, the toroidal field, $B_T$, normalization is used for consistency with the International Tokamak Physics Activity (ITPA) magnetohydrodynamics (MHD) topical group's joint experiment "MDC-19:  Error Field Control at Low Plasma Rotation" scaling \cite{Logan2020RobustnessScaling}.

\begin{table*}[h]
  \centering
  \def\arraystretch{1.5}
  \begin{tabular}{|l|l|c|c|c|c|c|}
    \hline
    Data & Fit & $\alpha_c$ & $\alpha_n$ & $\alpha_B$ & $\alpha_R$ & $\alpha_\beta$  \\
    \hline
O,L   & WLS &  $-3.46\pm0.05$  &  $+0.64\pm0.06$  &  $-1.14\pm0.08$  &  $+0.20\pm0.07$  &  $+0.15\pm0.07$ \\
O,L,H       & WLS &  $-3.62\pm0.04$  &  $+0.53\pm0.06$  &  $-0.95\pm0.07$  &  $+0.14\pm0.08$  &  $-0.19\pm0.05$ \\
    \hline
  \end{tabular}
  \caption{Scaling exponents from kernel density estimate weighted least squares (WLS) regressions for $n=1$ dominant mode overlap EF thresholds used to project the threshold to SPARC.}
  \label{tab:Exponents}
\end{table*}

\subsection{\label{sec:nScaling} n=1 Error Field Scalings}

The ITPA MDC efforts have produced a multi-machine database from which empirical scaling laws are constructed to project the EF thresholds in tokamak plasmas. The idea behind these scaling laws is, like the popular confinement scaling laws \cite{Goldston1984EnergyHeating}, to provide the macroscopic dependencies across wide ranges of parameter space rather than to capture every detail of the local behaviors in each device or scenario. 
With the overlap parameter, ideal MHD plasma response physics is used to capture the ideal outer layer response away from the resonant surfaces and the empirical scalings are used to represent the more complex resistive inner layer dynamics at the rational surface. Such scalings have provided predictions for the threshold in the ITER device before its operation, influencing its strict construction tolerances and EF correction coil designs \cite{Buttery2012LimitsITER,Amoskov2005StatisticalCorrection, Amoskov2019AssessmentBuilding, Lazerson2015}. 

The development of such a broad, multi-machine EF scaling law comes with a number of inherent challenges. A small set of scalar quantities are used for the scalings, motivated by (1)~lack of profile data in many of the discharges, (2)~applicability for future machine design where profile details are uncertain, (3)~unit-based checks like Connor-Taylor invariance \cite{Connor1977ScalingConfinement, Connor1988InvarianceConfinement} and (4)~database covariance checks to ensure good condition number fits. The historical variation in EF scaling experiments means that some parameter spaces are more highly sampled than others, and a number of weighting schemes have been investigated to compensate for this \cite{Logan2020RobustnessScaling}. 

The exponents described in Table 1 of Ref. \cite{Logan2020RobustnessScaling} describe empirical fits to a power law scaling for the critical overlap of the form,
\begin{equation}\label{eq:scaling}
    \delta_{c} = 10^{\alpha_c} n_e^{\alpha_n} B_T^{\alpha_B}
    R_0^{\alpha_R} \left(\beta_N/\ell_i\right)^{\alpha_\beta}.
\end{equation}
Here, and throughout this work, the line average density $n_e$ has units of $10^{19}$ m$^{-3}$, the on-axis toroidal field $B_{T}$ has units of Tesla, the major radius $R_0$ has units of meters and the normalized pressure $\beta_N / \ell_i$ is unitless. This critical overlap is the EF threshold at which resonant fields are expected to penetrate in the core, causing unacceptable locked modes.

\begin{table}[h]
  \centering
  \begin{tabular}{|l|l|l|c|}
    \hline
    Data & Fit & Scenario & Threshold ($\delta_c$)  \\
    \hline
O,L         & WLS &  SPARC L-mode & $1.1e-4$ \\
O,L,H       & WLS &  SPARC H-mode & $1.4e-4$ \\
O,L,H       & WLS &  ITER Baseline & $1.9e-4$ \\
    \hline
  \end{tabular}
  \caption{Nominal projections of the EF threshold to SPARC and ITER using the scalings from Table \ref{tab:Exponents} and scenario parameters from Table \ref{tab:equil_scalars}.}
  \label{tab:Thresholds}
\end{table}

Of the 6 empirical fits offered in Ref. \cite{Logan2020RobustnessScaling}, the weighted least squares (WLS) and down-sampled ordinary least squares (DSOLS) predict a nominal SPARC L-mode threshold of $1.1e-4$ while the largest prediction is $2.0e-4$ from the ordinary least squares fit that includes H-mode data. The lowest SPARC H-mode nominal threshold is $1.2e-4$ (DSOLS with H-mode data) and the highest is $2.3e-4$ (DSOLS without H-mode data). The WLS fit including H-mode data predicts a middle-ground SPARC H-mode nominal threshold of $1.4e-4$. This WLS fit weights the unique high field data from Alcator C-Mod more heavily, which we take to be a good approach in projecting to SPARC. Thus, for the remainder of this work we use the maximally conservative WLS fit of the Ohmic and L-mode data for projecting the EF threshold to the SPARC L-mode and the same WLS fit technique including H-mode data when projecting to the SPARC H-mode. The corresponding exponents are reproduced in Table \ref{tab:Exponents}. The corresponding EF threshold projections to SPARC and ITER are reported in Table \ref{tab:Thresholds}.

\begin{figure}[h]
\centering{}\includegraphics[width=1\linewidth]{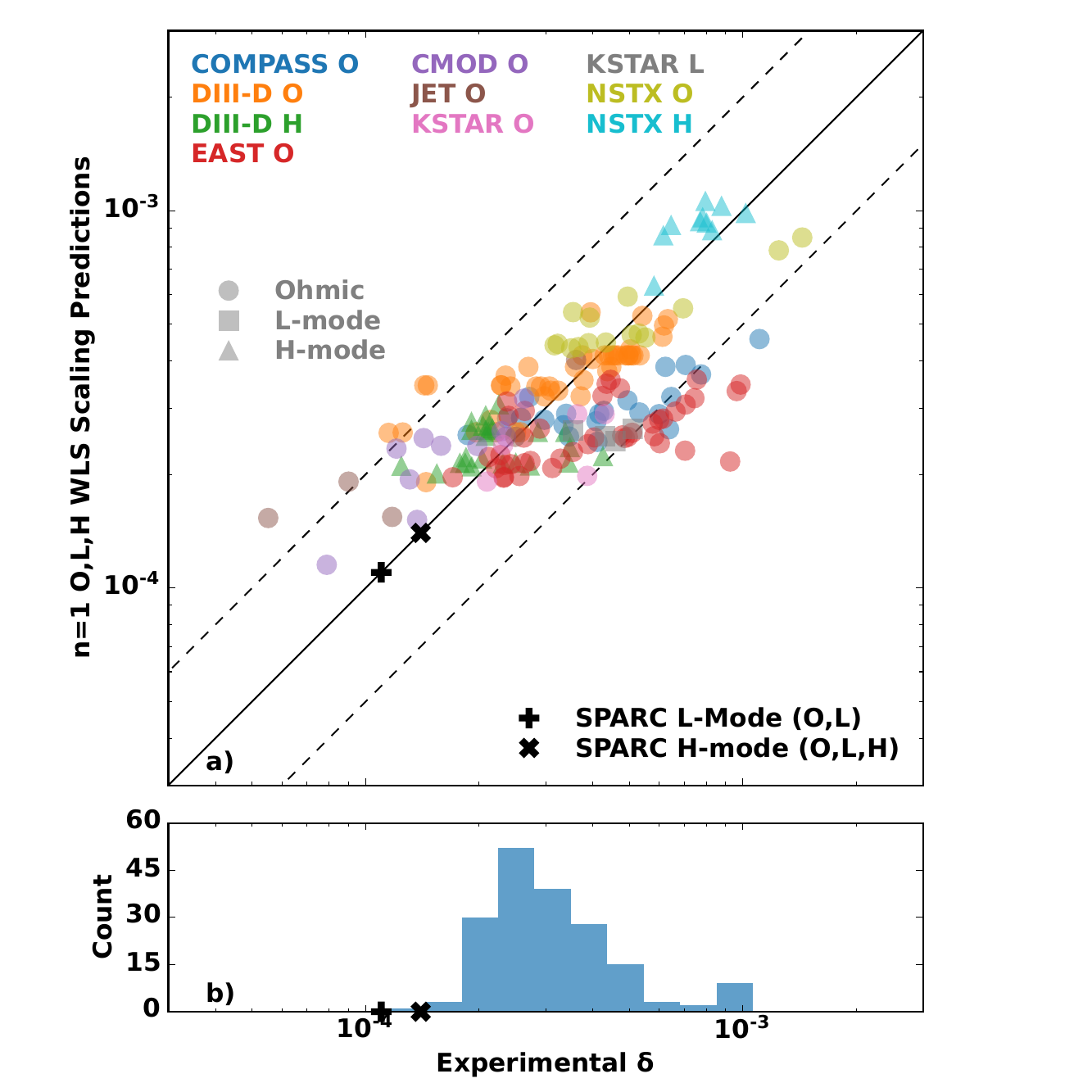}\protect\caption{(a) The weighted least squares ITPA EF threshold scaling plotted against the actual data in the ohmic, L-mode and H-mode database using in the fit. (b) Histogram of the empirical threshold data with SPARC scenario projections marked. Note the SPARC L-mode projection uses a scaling law fit to only Ohmic and L-mode data. \label{fig:scaling_vs_data}}
\end{figure}

We include ITER in these tables for comparison. The SPARC scenarios are projected to be more sensitive than ITER, but are within a factor of 2 of the ITER threshold. The SPARC nominal projections are within the experimental range measured in existing device scenarios, but near the lower end. This can be seen in \fref{fig:scaling_vs_data}, where the SPARC projections lie in the lower tail of the histogram of existing empirical thresholds. 

\begin{figure}[h]
\centering{}\includegraphics[width=1\linewidth]{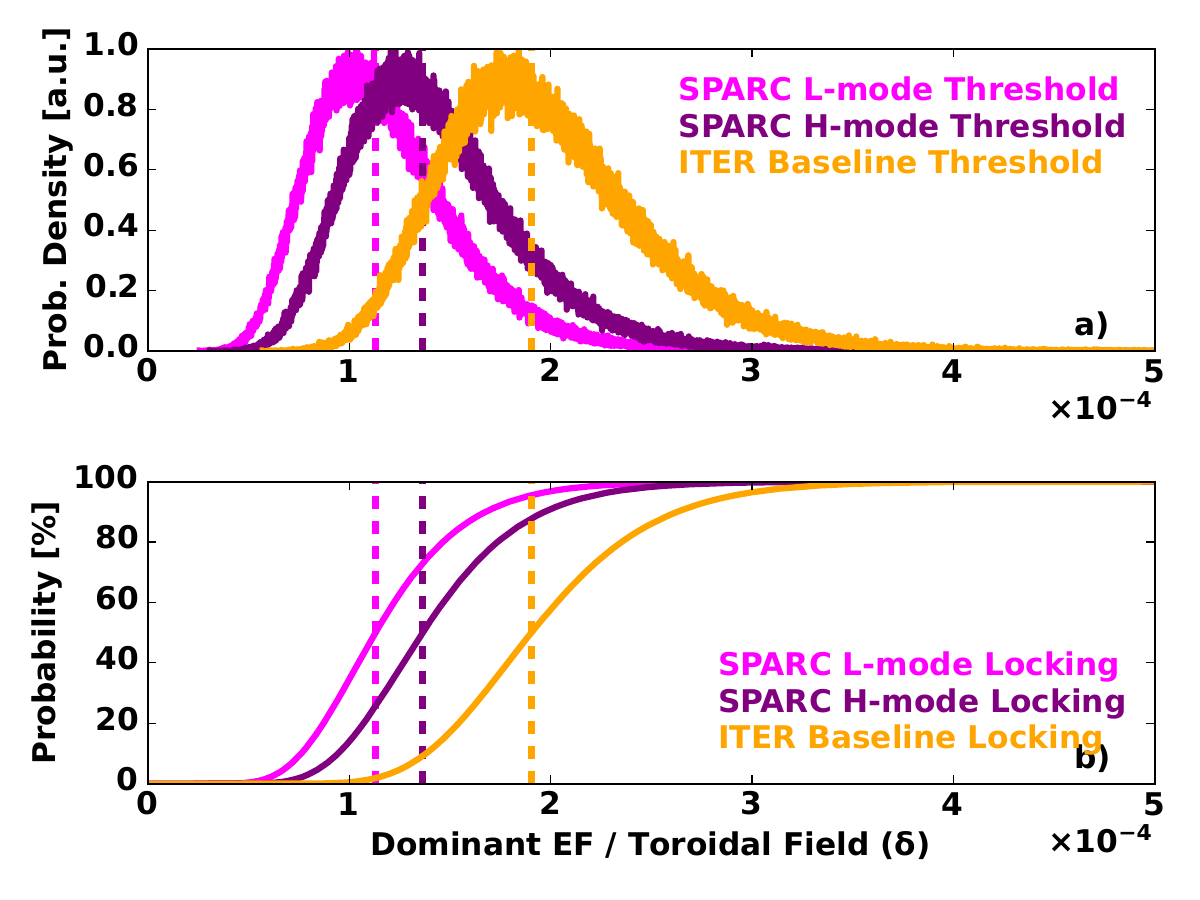}\protect\caption{(a) Probability distribution function of dominant mode overlap thresholds and (b) corresponding risk of locking for a given EF value in the scenarios defined in Table \ref{tab:equil_scalars}. Dashed vertical lines mark the threshold using the nominal value of each exponent in the corresponding scaling law. \label{fig:scaling_pdfs}}
\end{figure}

It is important to note that there is a significant uncertainty in these projections. 
This can be seen in the large scatter of the scaling thresholds plotted against the underlying data in \fref{fig:scaling_vs_data}, where dashed lines mark a factor of 2 variation from the diagonal.
The relatively large uncertainties in the exponents reported in Table \ref{tab:Exponents} mean that simple Taylor-Expansion-based uncertainty propagation is not accurate here. Instead, we adopt a probabilistic approach to describing the EF threshold and risk of locking. In this approach, we randomly sample the exponents from a normal distribution defined by the nominal value and standard deviation reported in Table \ref{tab:Exponents} to build a numerical model of the threshold probability distribution function (PDF). The integral, or cumulative distribution function, then gives the projected risk of locking for a given EF overlap value. These distributions are shown in \fref{fig:scaling_pdfs}.

\subsection{\label{sec:nonresonant} Nonresonant Error Fields}

The dominant mode quantification of risk is a powerful model for identifying and correcting the most deleterious component of the EF spectrum that has proven highly successful in guiding EFC in current devices. It is important, however, to recognize the known limitations of such a model. There are, for example, a number of ways for the EF to impact stability and performance without any dominant mode overlap. These include,
\begin{itemize}
    \item The drive of $n=1$ core resonant fields through subdominant core-coupling RSVs (in the case that the core region contains multiple $n=1$ rationals),
    \item The drive of edge resonant $n=1$ fields,
    \item The creation of neoclassical toroidal viscosity (NTV) from non-resonant $n=1$ EFs  
    \item $n>1$ EFs driving core resonant fields, edge resonant fields, and/or NTV.
\end{itemize}

We will address the first three of these in turn. The last, we encompass through a conservative posture as to how effective we project any $n=1$ EFC can be but do not explicitly treat here. The ways in which future work could explicitly address $n>1$ physics will be discussed in Sec. \ref{sec:Summary}.

\subsubsection{Subdominant $n=1$ Core-Coupling}

The first of the limitations discussed in Sec.~\ref{sec:nonresonant} is trivially avoided in the SPARC H-mode, which only has one $n=1$, $m>1$\footnote{Note, any $m,n = 1,1$ surface is always excluded from the perturbative MHD GPEC calculations as it naturally produces an large internal kink mode in the model but is avoided by sawteeth in reality.} rational within the core of the plasma as defined by the ITPA database. The L-mode reference equilibrium has two surfaces, and thus two RSVs of the resonant coupling matrix. The singular values of these are 0.57 and 0.45 respectively, making the first not decisively "dominant" over the other. The second mode is a midplane inboard weighted structure with a poloidal half wavelength of $\sim0.8$ m. Thus, EFs from the central solenoid (CS) coils might be expected to drive core resonant fields through this mode while EFs from the outboard poloidal field (PF) coils would likely not. It is known that the largest, coherent asymmetries from the entire central column effectively shift the nominal plasma reference frame rather than driving tearing \cite{Halpern2025DeterminingEquilibria} but there is still a finite CS drive expected from leads, winding asymmetries, and incoherent assembly location errors.  
Just from the singular values, however, we can already begin to see that secondary effects might have a $\sim44$\% contribution to the EF sensitivities. If the intrinsic EF were to overlap with both modes equally, we might expect locking if the EF exceeds $\sim2.3 \delta_c$ even if perfectly correcting the dominant mode.

\subsubsection{Edge Resonant $n=1$ Fields}\label{sec:edge_resonnaces}

Edge resonant $n=1$ EFs are known to be important for access to H-mode \cite{Yang2025InfluenceField} and for control of edge localized mode (ELM) stability \cite{Yang2020LocalizingKSTAR, Kim2024HighestTokamak}. A full quantitative assessment of the impact on the edge stability and transport of edge EFs is beyond the scope of the current paper, but an important aspect of H-mode tokamak design in its own right. We note the higher $q$ edge rationals are sensitive to higher $m$ perturbations that have a faster radial decay from their source to the plasma. Similarly, the higher $n$ perturbations are known to be important for H-mode access \cite{Schmitz2019LHPerturbations, Kriete2020InfluenceTokamak} as well as pedestal stability and performance \cite{Kim2024HighestTokamak} but decay faster between the EF sources (PF coils, CS coils, etc.) and the plasma \cite{Logan2020PhysicsTokamaks}. 

\begin{figure}[h]
\centering{}\includegraphics[width=1\linewidth]{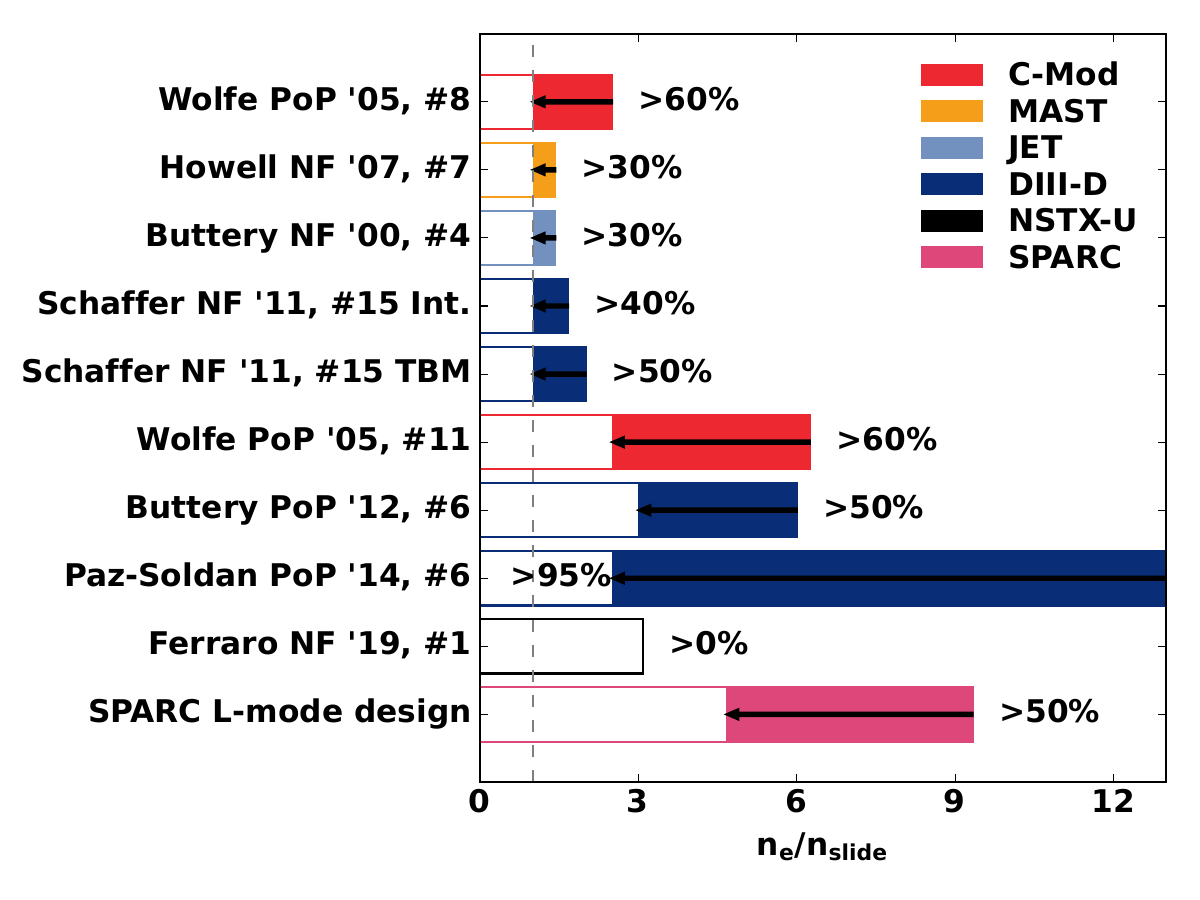}\protect\caption{The lowest density accessible with EFC (open bar) and without EFC (closed bar) for a number of published EFC studies. The densities have been normalized to the density at which REs are expected (which was directly determined for DIII-D in \cite{Paz-Soldan2016TheLimit}). As such, no amount of EFC can enable access below 1. \label{fig:EF_limits_density}}
\end{figure}

To assess the impact of the smaller wavelength (higher harmonic) EFs, we turn to the experiences reported in existing literature on the limits of EF correction. The $n>1$ residuals did not seem to have a significant effect when correcting broad-spectrum EFs from a test blanket module (TBM) in DIII-D Ohmic plasmas \cite{Schaffer2011}, as seen in the ability to obtain similar low density limits when correcting the TBM as where obtained when correcting the intrinsic EF (Fig. 15 of \cite{Schaffer2011}). This metric, the lowest density obtainable, has been used often to check the quality of EFC. A summary of such tests is given in \fref{fig:EF_limits_density}. Unfortunately, these tests are complicated by the fact that the low density plasmas terminate due to runaway electron (RE) effects \cite{Paz-Soldan2016TheLimit} that are not the concern of this study (due to the plan to operate SPARC well above this limit, $n_{slide}$). Even perfect axisymmetric plasmas would not be able to access $n_e/n_{slide} < 1$. Still, this summary shows it is rare that EFC experiments are able to reduce the density limit by significantly more than 50\%. As the same wave of single-machine EF studies reported roughly linear density scalings (presumably different from the full database scaling due to the correlations between density and rotation or temperature in a given experiment), this corresponds to roughly 50\% effective EFC being a reasonable expectation from past experience. This is the expectation we adopt for SPARC, as indicated on the final bar of figure \fref{fig:EF_limits_density}. 

\subsubsection{Non-resonant Rotation Braking}

The creation of NTV through the non-resonant EF spectral components is conflated with the higher mode number residual EF effects in \fref{fig:EF_limits_density}. The DIII-D TBM studies found that either 30\% or 75\%, depending on the plasma pressure, of the rotation reduction due to the TBM EF could be recovered with EFC. Other DIII-D studies showed rotation was much more severely degraded, compared to density or pressure, by the residual EF after correcting the dominant mode overlap of one 3D coil array with another \cite{Paz-Soldan2014TheImportanceDIII-D, Paz-Soldan2015DecoupledFields}. These rotation breaking observations are thought to be due to the NTV torque caused from the residual EF spectrum. This NTV is driven even if there are no sub-dominant resonant modes. Only a special subset of the EF spectrum, analogous to quasi-symmetry, does not cause such braking \cite{Park2021QuasisymmetricTokamaks}. 

To quantify the expected impact of residual EF braking effects on SPARC EFC efficacy, we introduce here a simplified model of the NTV effects. 
First, without NTV effects, the condition describing the required current ($I$) in any given EFC coil set is,

\begin{equation} \label{eq:EFC_model_noNTV}
    \delta_\mathrm{c} > \delta_\mathrm{EF} - C I.
\end{equation}

Here, $C$ is the linear coupling between the EFCCs and the dominant mode. According to tearing theory \cite{Fitzpatrick2012NonlinearPlasmas, Hu2020NonlinearThreshold} and empirical findings \cite{Buttery2000ErrorJET, Lazzaro2002ErrorSpin-up}, the resonant field at which locked modes appear is proportional to the plasma rotation. Thus, we modify Eq.~\eqref{eq:EFC_model_noNTV} to account for the impact of the EFCC on the plasma rotation,

\begin{equation} \label{eq:EFC_model_NTV}
    \delta_\mathrm{c}\left[\frac{T_0 - T_{NR} I^2}{T_0}\right] > \delta_\mathrm{EF} - C I.
\end{equation}

This model assumes steady state torque balance with a constant momentum confinement time such that, $dL/dt = 0 = T_0 - T_{NR}I^2 - L / \tau_L$, where the angular momentum $L$ is proportional to the toroidal rotation.
The factor in square brackets then makes the critical overlap linearly proportional to the new rotation with EFCC current $I$ applied.
Here, $I$ is amplitude of $n=1$ current in a given EFCC setup (fixed phasing if multiple arrays are used) in units of kA. 
$T_0$ is the initial torque spinning the plasma and $T_{NR}$ is the NTV braking torque per kA$^2$ of the EFCC field spectrum with the dominant mode component subtracted. 
The dominant mode component does not contribute to the braking because it is assumed to be canceling the intrinsic EF. 
This makes the conservative assumption that the EFC is always adding to the intrinsic NREF, which is left as an unknown. 

If one aims to fully correct the intrinsic EF, the maximum correctable intrinsic EF would be simply given by the point at which the rotation becomes 0 and the plasma is susceptible to locking from any infinitesimal perturbation, 

\begin{equation} \label{eq:EFC_model_NTV_torque_balance}
   T_0 - T_{NR} (I_{max})^2 = 0,
\end{equation}
which results in the constraint,

\begin{equation} \label{eq:EFC_model_NTV2}
   \delta_{max} = C I_{max} = C \sqrt{\frac{T_0}{T_{NR}}}. 
\end{equation}

Note that we have introduced negative signs on the torque coefficients here so that $T_{NR,R} > 0$ corresponds to NTV rotation braking. The model assumes the $\mathbf{E}$$\times$$\mathbf{B}$ rotation in the core of the plasma is above zero in the electron diamagnetic direction and that the ion NTV is dominant such that the NTV brakes the plasma towards a negative neoclassical offset. If the rotation is below zero, NTV drive of the rotation towards a finite offset would have to be taken into account and would actually be stabilizing for initial rotations between the offset and zero \cite{Garofalo2008ObservationTokamak}. 

\begin{figure}[t]
\centering{}\includegraphics[width=1\linewidth]{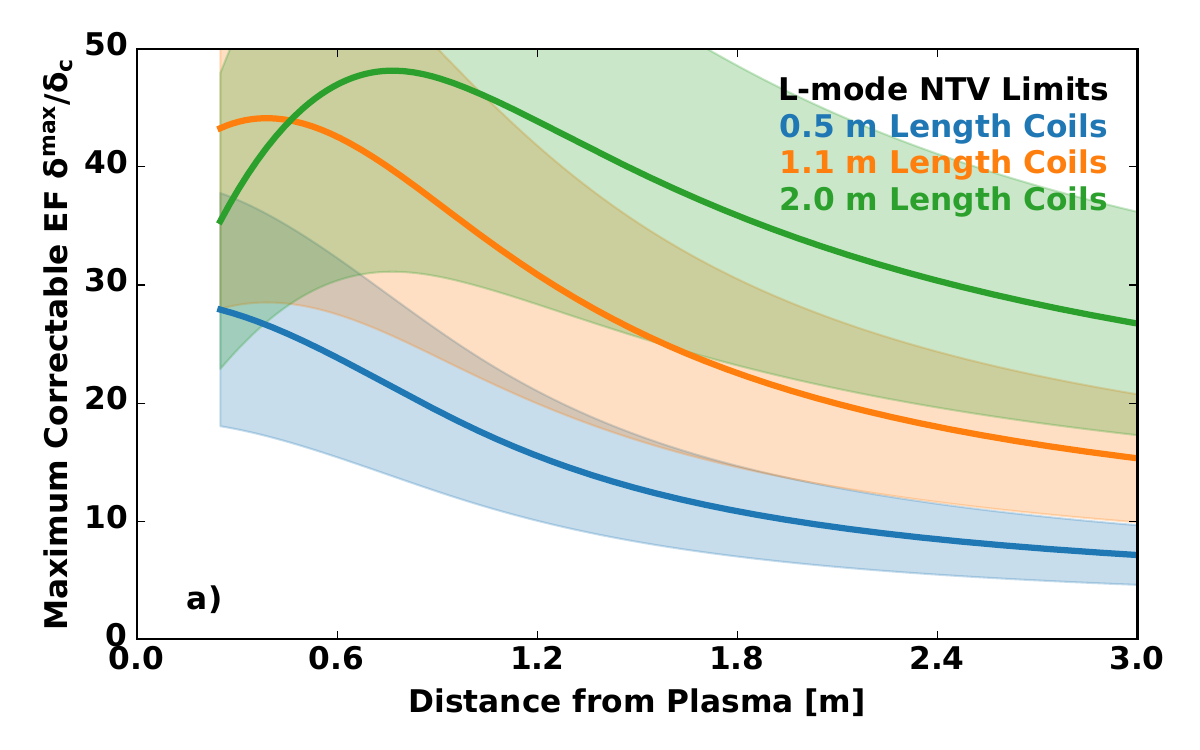}
\includegraphics[width=1\linewidth]{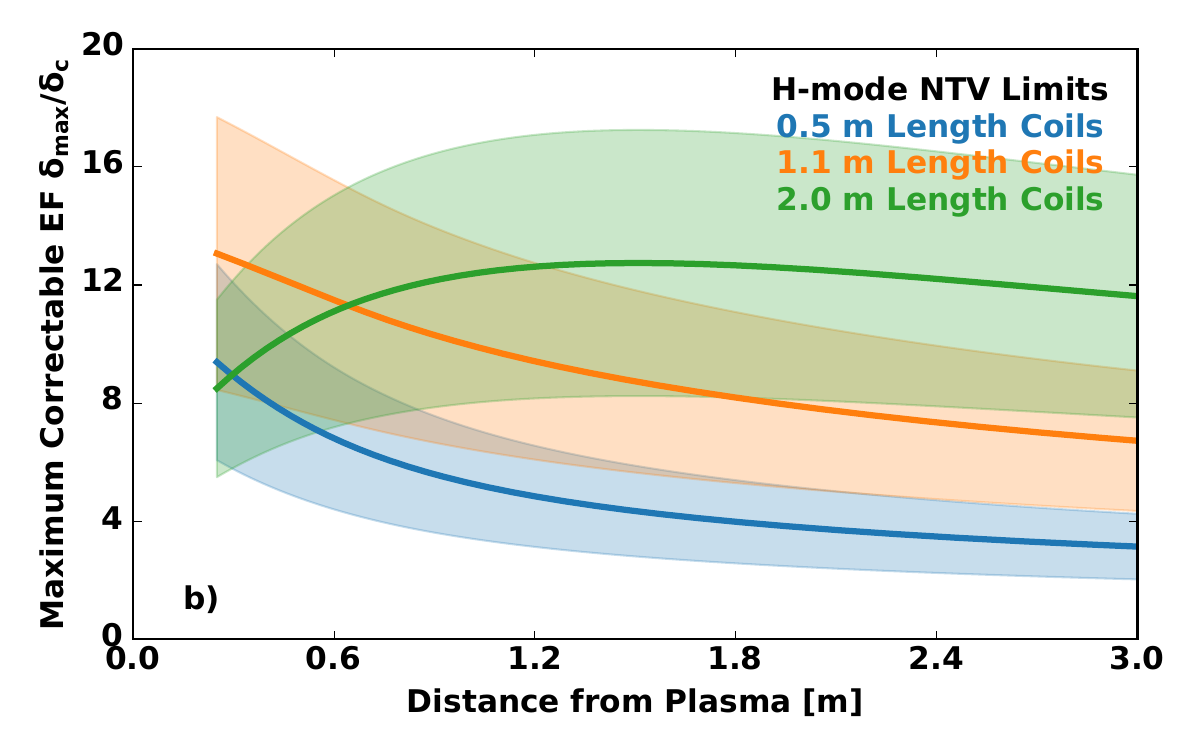}
\protect\caption{The maximum correctable EF normalized by the nominal threshold value for midplane, vertical coil arrays of 3 different heights as a function of the radial distance the array is from the L-mode (a) and H-mode (b) plasmas. \label{fig:ntv_limits_lmode}}
\end{figure}

Realistic coils inevitably do not match the dominant mode spectrum perfectly, but the matching can vary greatly within reasonable design constraints \cite{Logan2020PhysicsTokamaks}. 
Figure \ref{fig:ntv_limits_lmode} shows how the non-resonant spectral pollution of midplane (i.e. centered on $z=0$ like the M EFCCs are) arrays limits the maximum correctable error field in the L- and H-mode target plasmas.
The correctable error field is normalized by the nominal threshold from the Table \ref{tab:Thresholds}, such that $\delta_{max}/\delta_c=2$ on the y-axis would correspond to the stance taken in Sec.~\ref{sec:edge_resonnaces}. 
As is appropriate, the NTV limits are projected to be above 2 for reasonable coils.
To acknowledge the high degree of uncertainty in NTV modeling and the projection of intrinsic torques (which in turn, influence the NTV), we've included uncertainty bands in the figure corresponding to $\pm$50\% uncertainty in $T_{NR}$ and $T_0 = 4\pm2$ N$\cdot$m. These, if anything, are perhaps optimistic uncertainty estimations. 
The takeaway from this exercise is that NTV has the ability to significantly impact the correctable error field for reasonable coil array sizes in practical locations. 
The NTV is less limiting in L-mode where the lower plasma amplification of EFs and lower rotation result in less NTV torque. 
However, it still limits operable construction EFs to within an order of magnitude of the natural locking threshold.
The choice to not assume EFC efficiency above a factor of 2 is below the expected NTV limits for reasonable coils of interest in both plasmas, leaving room to allow for accumulation of sub-dominant resonances, edge EFs, and NTV effects.

\subsection{\label{sec:n1_EF} $n=1$ Error Field Model}

Common sources of error fields in tokamaks include as-designed coil asymmetries such as turn windings and lead junctions as well as deformations and displacements from the nominal design that occur during construction and assembly of the primary coils. This section details the EFs from one iteration of the nominal designs for the SPARC PF and CS coils (not necessarily the final design) and outlines a system for assessing the locked mode risk given a nominal design and construction tolerances.
The examples here concentrate on toroidally encompassing current paths, including the jumper paths between TF coils for example. Details of the toroidal field (TF) coil windings are not considered as they have a natural $n\gg1$ symmetry. 
Coherent $n=1$ tilts and shifts of the TF system shift the nominal axis of symmetry about which linear MHD calculations should perturb \cite{Halpern2025DeterminingEquilibria}. We thus take the TF coils as our reference coordinate system, representing any $n=1$ TF displacement in the lab frame as an equal and opposite PF+CS coil shift in our TF frame.

\begin{figure}[h]
\centering{}\includegraphics[width=1\linewidth]{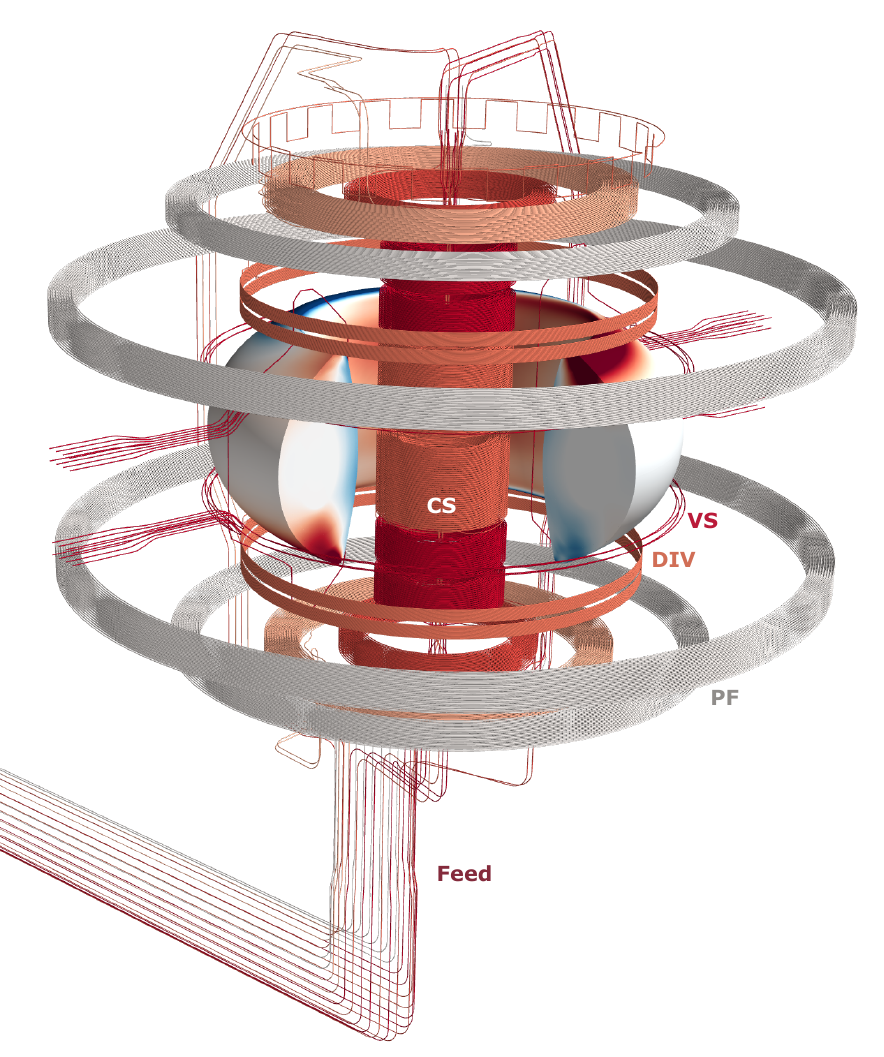}\protect\caption{SPARC Poloidal Field (PF), Central Solenoid (CS), Divertor (DIV), and vertical stability (VS) coil windings with their junctions and feed lines. Also shown are the TF coil jumpers and vacuum vessel seam lines. The plasma surface shows the H-mode reference $n=1$ EF from all the nominal sources.
\label{fig:coil_system}}
\end{figure}

The PF and CS coils are nominally circular coils centered on the tokamak central axis at major radius $R=0$. 
These coils are comprised of many turns of current conducting tape, and the winding of these turns introduces small toroidal symmetries to the coils. 
Additionally, the junction between the coil winding start/end points and the mostly-self-canceling in/out feed system introduces an inevitable asymmetry. 
A detailed documentation of the windings and junctions is outside the scope of this manuscript, but a overview of the many coils with examples of these sources of asymmetry called out is shown in \fref{fig:coil_system}.

\begin{figure}[t]
\centering{}\includegraphics[width=1.0\linewidth]{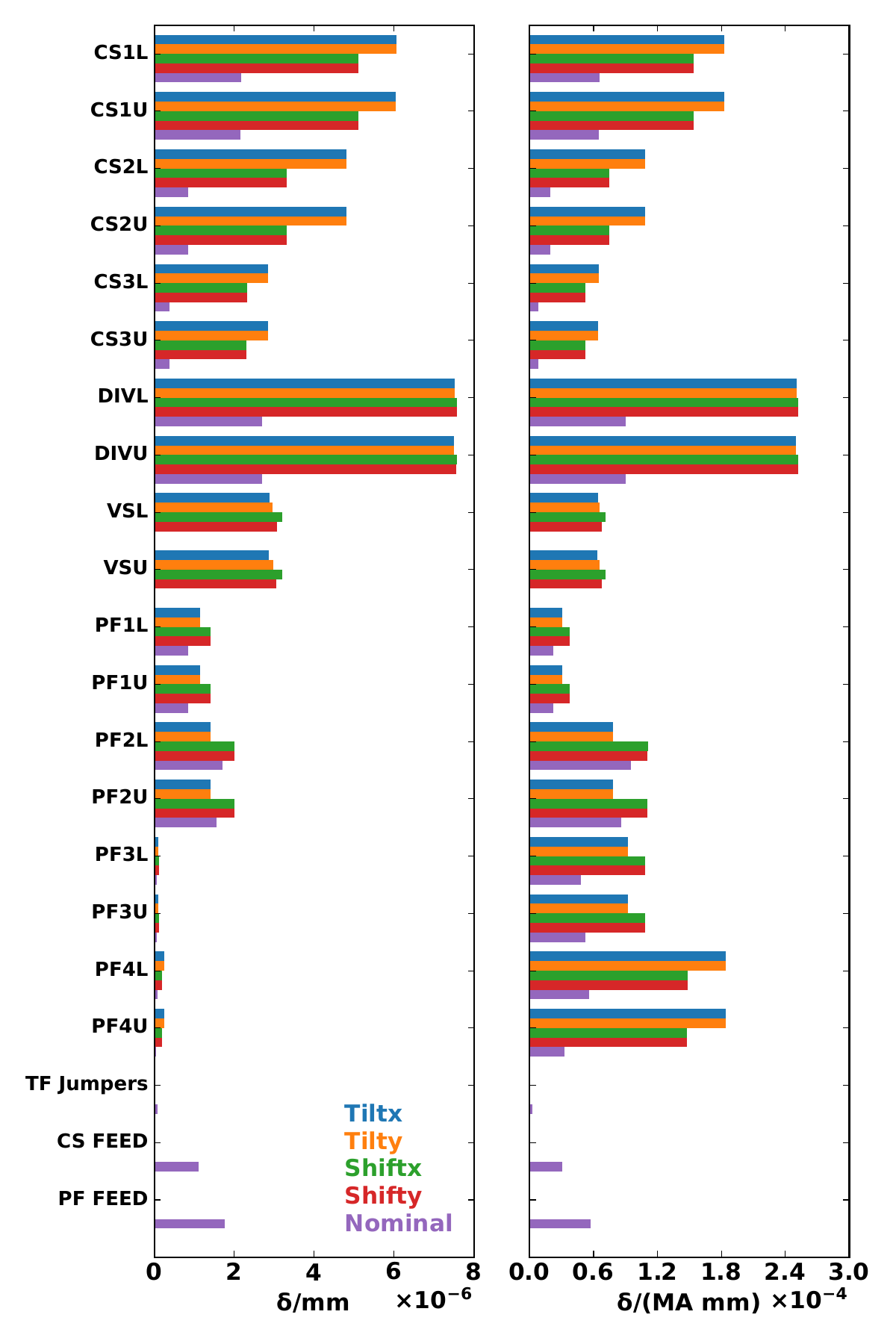}\protect\caption{Bar chart of L-mode overlap per mm of shift/tilt of each coil and overlap per mm per MA. The nominal overlap's in each do not include any mm normalization, and are included as a reference to assess the importance of assembly accuracy compared to winding design. \label{fig:EF_per_mm_L}}
\end{figure}

\begin{figure}[t]
\centering{}\includegraphics[width=1.0\linewidth]{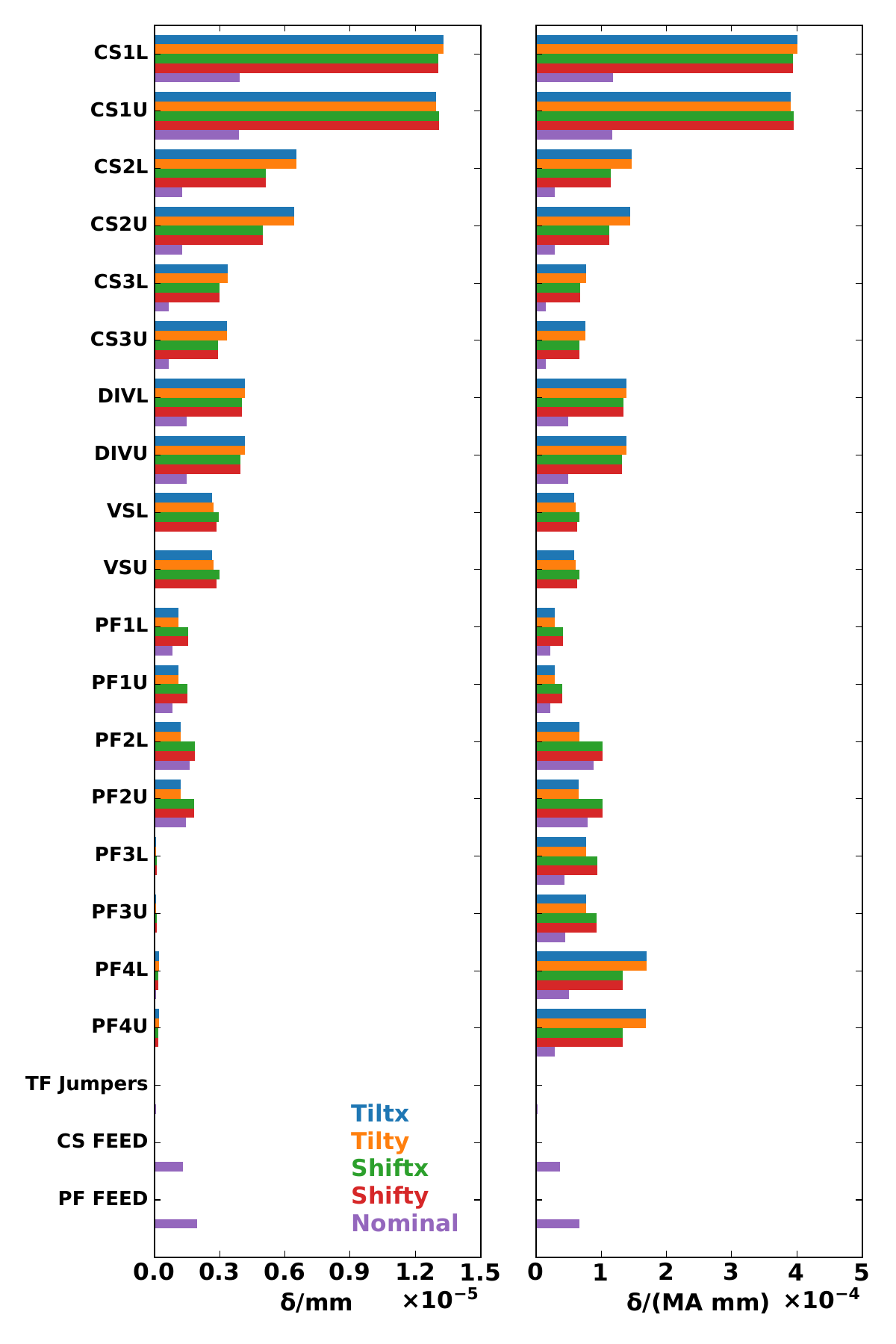}\protect\caption{Bar chart of H-mode overlap per mm of shift/tilt of each coil and overlap per mm per MA. The nominal overlap's in each do not include any mm normalization, and are included as a reference to assess the importance of assembly accuracy compared to winding design. \label{fig:EF_per_mm_H}}
\end{figure}

Given the highly accurate CAD based current paths in the windings, feed junctions, and feed systems themselves, the $n=1$ EF from the appropriate current in these coils for each reference discharge can be immediately calculated in GPEC. 
The individual contributions of each independent source considered are given in figures \ref{fig:EF_per_mm_L} and \ref{fig:EF_per_mm_H}.
The naming convention uses "U" for upper or "L" for lower coils and counts from the midplane, so the central solenoid coils from bottom up are [CS3L, CS2L, CS1L, CS1U, CS2U, CS3U].  
Coil-feed "junctions" are included with the coil windings such that the "feed" lines do not include any divergence of the in and outgoing lines to the spatially separated winding start and endpoints.
The sources considered include CS, PF coils, and the "jumper" feed lines to/from/between TFs. 
Some destructive phasing results in total error fields well below the sum of source amplitudes. 
The total nominal design EF overlap $\delta_{nom}$ is 5.45e-6 for the L-mode and 1.77e-5 for the H-mode, which are each an order of magnitude or more below the critical values given in Table \ref{tab:Thresholds}. 

The $n=1$ fields from small displacements of the nominal coil placement, however, can create larger EFs. 
Figures \ref{fig:EF_per_mm_L} and \ref{fig:EF_per_mm_H} show the $\delta$ per mm of shift or tilt in a given coil next to each coil's nominal design EF. 
The figure shows that small shifts and tilts of the coils upon installation, even just one mm, can quickly dominate over the as-designed EF from turn windings and feed junctions. 
This is not an inevitable truth but is rather the result of care taken in designing the coils themselves to have minimal $n=1$ asymmetry. Engineering designs of pancake transition lengths and phasing optimizations, lead positioning, and other asymmetries where made with EF minimization in mind at each step in the process.

Here, shifts are rigid displacements of the entire coil winding in the $x$, $y$, or $z$ direction such that the new coil has points ($x+\Delta x$, $y+\Delta y$, $z+\Delta z$). Shifts are reported throughout this work in mm.
Tilts are a tilt of planar coils about an in-plane axis ($x$ or $y$). Tilts are also reported in mm.  Adopting $x,y,z$ coordinates from the coil’s center of mass ($x_0,y_0,z_0$) is convenient for describing tilts. The angular distortion amplitude $\gamma_y$ is calculated from a mm tilt amplitude $A_y$ using $\gamma_y = \arcsin(A_y / R_{nom})$. Here, $R_{nom}$ is the nominal radius of the coil calculated from a center of mass calculation of its windings. The coil winding node points are then perturbed using,
\begin{align}\label{eq:tilty}
    R &= \sqrt{x^2 + y^2}, \\
    \alpha  &= \mathrm{atan2}(z,x) + \gamma_y,\\
    \Delta x &= (R * \cos(\alpha) - x),\\
    \Delta z &= (R * \sin(\alpha) - z).
\end{align}

\begin{figure}[t]
\centering{}\includegraphics[width=1\linewidth]{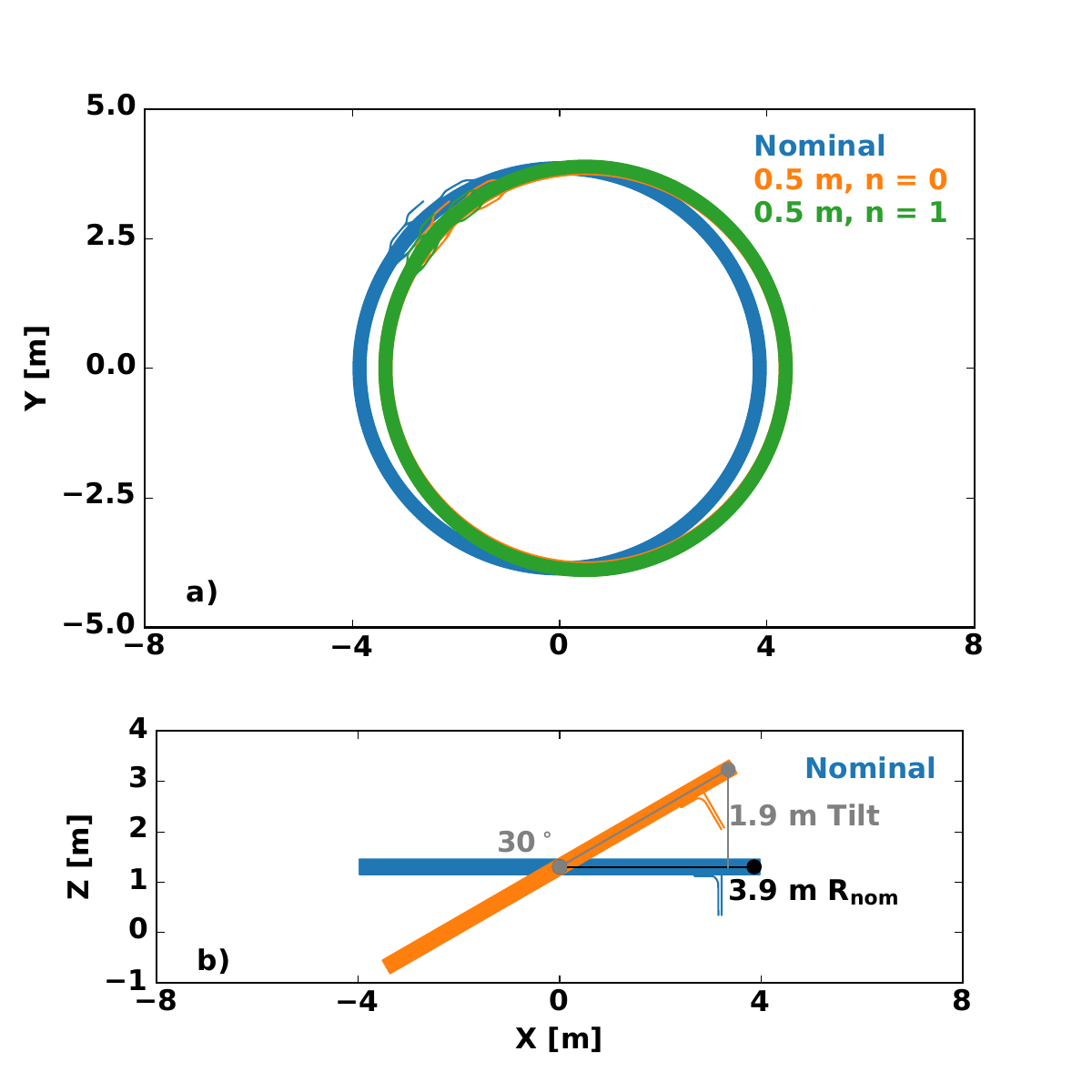}\protect\caption{Extreme examples of a 500 mm x-directed shift (a) and 30 degree y-axis tilt (b) for the upper-midplane PF coil (PF4U).
\label{fig:shift_tilt}}
\end{figure}

\noindent Similar calculations are done for the perturbations around the x axis, which results in perturbations $\Delta y$ and $\Delta z$.
Note, tilts can be generalized to higher $n$ using  $\alpha = \mathrm{atan2}(z,x) + \gamma_y \cos\left((n-1)*\phi\right)$. The $n=2$ perturbation then corresponds to a saddling of the coils.
A similar extension of shifts to arbitrary $n$ is,
\begin{align}\label{eq:petal}
    \Delta_R &= A_x \cos(n\phi) + A_y \sin(n\phi), \\
    \Delta x &= \Delta_R * \cos(\phi),\\
    \Delta z &= \Delta_R * \sin(\phi).
\end{align}
This parameterization results in petal-like distortions. For small perturbation amplitudes, the $n=1$ petal distortion (i.e. beaning) is approximately a rigid shift. Figure \ref{fig:shift_tilt} shows this to be true even for a 500 mm amplitude for the upper-midplane PF coil (PF4U) in SPARC. Only rigid shifts (i.e. $n=0$ for x) are considered in this work. Considering both shifts and petals would be a nonphysical double-counting, as the installation positioning will be performed based on the true magnetic axis rather than the nominal circular coil. The $n=2$ petal distortion corresponds to (not elliptical) elongation. The $n=2$ distortions do not create impactful $n=1$ EFs, and their study is left to be reported in a separate work.

In addition to individual coil assembly tolerances, coherent tolerances have been implemented to account for multiple coils displacing together. 
In particular, the vertical stability coils have been given a 2 mm coherent shift tolerance to allow for possible displacement of these vacuum-vessel-mounted coils due to disruption forces. 
The CS coil stack is given a 0.5 mm coherent shift tolerance to account for movement of the full set of six CS magnets with respect to the magnetic axis of the TF inner legs when the TF system is energized. Such a shift between the TF and CS magnetic axes can develop due to slight variability in the thickness of the case material that is squeezed between each of the TF inner legs and the CS windings.
No perturbations are considered for the feed systems, including the TF jumper cables.

To assess the expected EF amplitude given a set of assembly and coherent displacement tolerances, a Monte Carlo method of randomizing shift and tilts is used to form a probability distribution of EFs.
To do this efficiently, the shift and tilt per unit of tolerance are calculated for every $i$th coil ($\delta^S_i, \delta^T_i$) with current corresponding to the appropriate scenario.
This approximated as independent of phase.
Figures \ref{fig:EF_per_mm_L} and \ref{fig:EF_per_mm_H} show this is a valid simplification as the x and y displacement EFs are approximately equal for the CS and PF coils.
The total EF in each Monte Carlo instance of the SPARC assembly is then calculated by applying a random amplitude and phase to each of the N coil shift and tilt,

\begin{equation}
    \delta_{\text{EF}} = \sum_i^C \left[ \delta^N_i + A^S_i \exp(\Theta^S_i) \delta^S_i + A^T_i \exp(\Theta^T_i) \delta^T_i \right],
\end{equation}

\noindent Here $\delta^N$ is the EF of the nominal (as designed) coil winding. $A^S$ and $A^T$ are random variables representing the magnitude of the shift or tilt relative to the tolerance. These are sampled from a weighted PDF $P_A(x) = 4x^3, x \in [0, 1]$ as was done in \cite{Pharr2024ErrorITER}. This models the physical reality of assembly, in which sub-tolerance adjustments will be made until a tolerance is met - making it more likely the final position will be near the edge of the tolerance. The phase of the shifts and tilts, $\Theta^S$ and $\Theta^T$ are random variables sampled from a uniform likelihood distribution $P_\Theta(x) = 1/(2\pi),  x \in [0, 2\pi)$. Note the $N$ coils include each individual coil as well as the coherent coil combinations. The coherent combinations have zero nominal EF so as not to double count. Any case with zero tolerance is assigned zero (not infinite) EF per tolerance.

\begin{figure}[t]
\centering{}\includegraphics[width=1.0\linewidth]{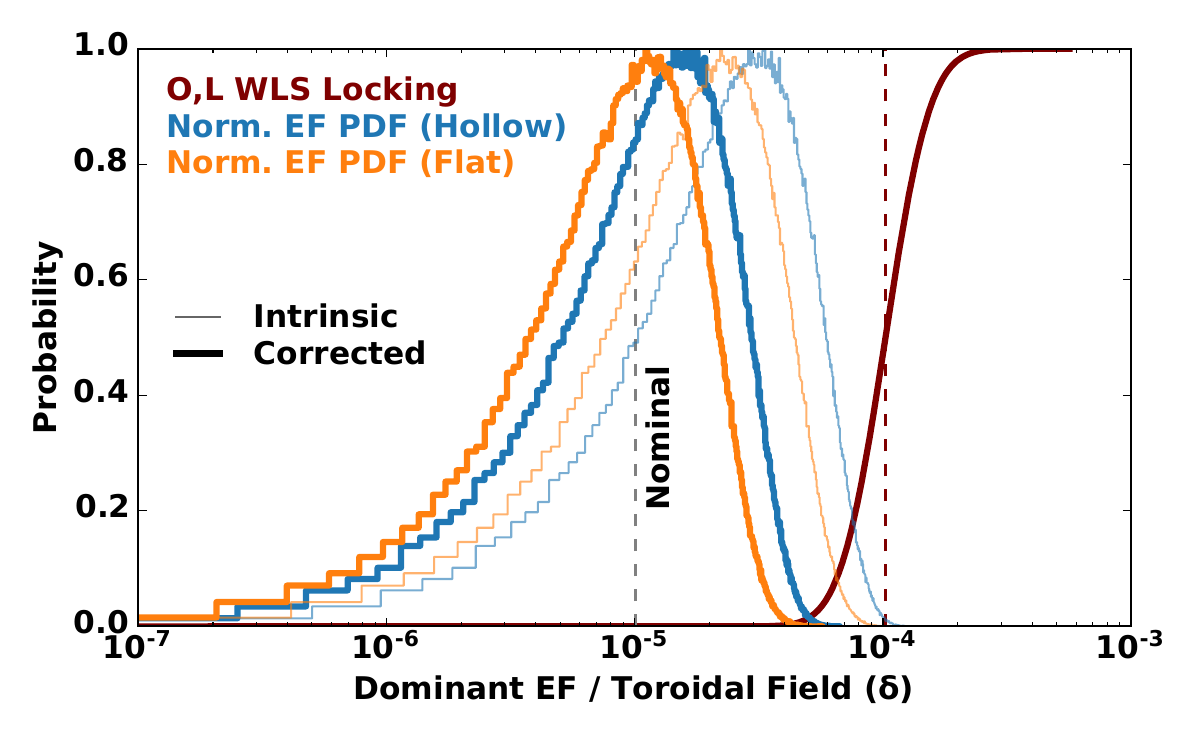}\protect\caption{Example Monte Carlo EF PDF and scaling threshold CDF for the SPARC L-mode. The overlap of these distributions determines the risk. The Monte Carlo intrinsic EF PDF is shown as a thin line for references, the PDF assuming a 50\% effective EF correction used for risk assessment is shown in bold. The nominal (as-designed) EF and nominal scaling law threshold are shown as dashed lines. \label{fig:EF_pdfs_L-mode}}
\end{figure}

\begin{figure}[t]
\centering{}\includegraphics[width=1.0\linewidth]{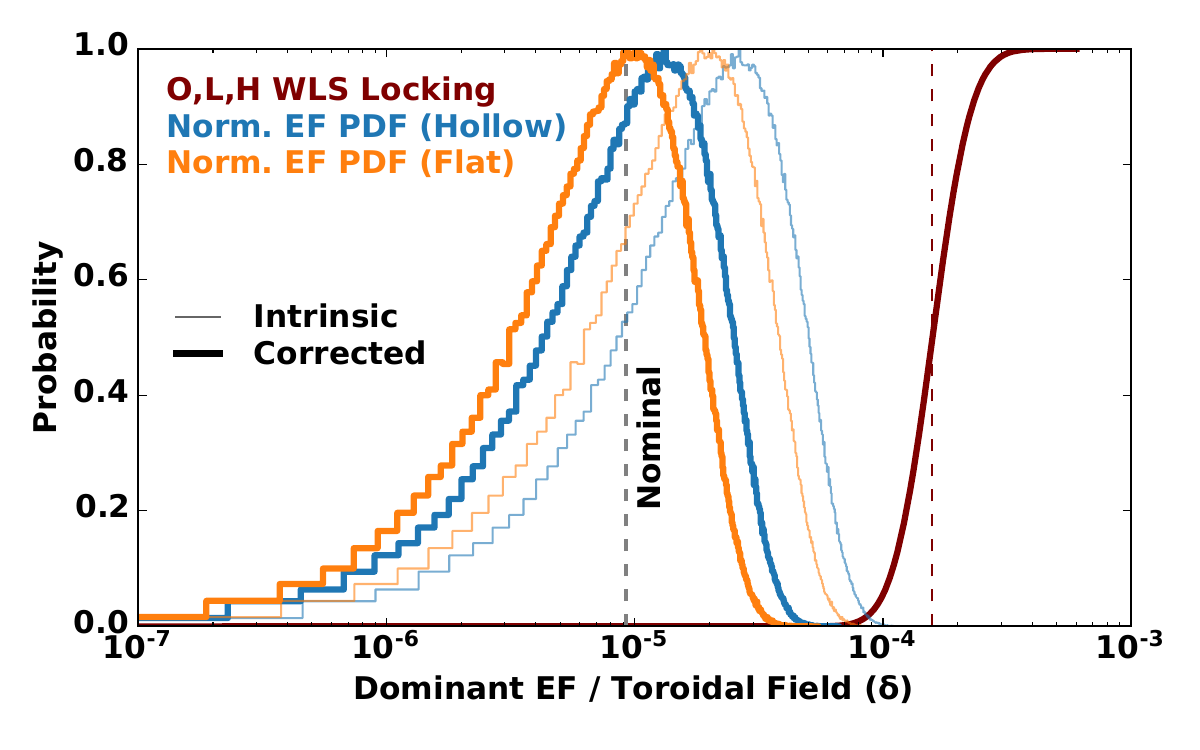}\protect\caption{EF PDF and scaling CDF as described in \fref{fig:EF_pdfs_L-mode} shown for the H-mode scenario. \label{fig:EF_pdfs_H-mode}}
\end{figure}

Example EF PDFs formed from $1e6$ Monte Carlo assembly instances and their corresponding EFs in the SPARC L-mode target are shown in \fref{fig:EF_pdfs_L-mode}. 
This is an example that uses a 1 mm tolerance for shifts and tilts of the CS coils and 3 mm tolerance for the PF, DIV, and VS coils in addition to the fixed coherent tolerances described above.
These are not the final assembly tolerances. They are only an example tolerance scheme that gives a sense of the order on which tolerances must be to maintain low risk.
The solid blue distribution uses hollow amplitude probabilities. 
The orange PDF shows, for comparison, the result if amplitudes were sampled from a uniform distribution. The choice of a hollow distribution intuitively increases the EF expectation value. 
The intrinsic EF Monte Carlo is shown as a thin line but the corrected EF PDFs in bold, which have the EF magnitude halved following the stance outlined in Sec. \ref{sec:nonresonant}, are the distributions used to assess operational risk. The same tolerances result in more separated distributions (less risk) in H-mode, as shown in \fref{fig:EF_pdfs_H-mode}.

\subsection{Risk Assessments}\label{sec:risk_assessment}

The risk of a given assembly tolerance scheme resulting in intrinsic EFs that lock the target scenario is given by both the probability of a EF magnitude manifesting and the cumulative probability that EFs of that magnitude or smaller lock the plasma. This total risk, $r$, can be written,

\begin{equation}
    r = \int_0^{\infty} E(\delta) \int_0^\delta L(\delta') d\delta' d\delta,
\end{equation}\label{eq:risk}

\noindent where $E(\delta)$ is the probability density function of the intrinsic EF shown in figures \ref{fig:EF_pdfs_L-mode} and \ref{fig:EF_pdfs_H-mode}, $L(\delta)$ is the PDF of the locking threshold shown in \fref{fig:scaling_pdfs} and the prime integral is the CDF shown in those figures. Note that this approach assigns a finite risk even when the nominal and expectation value EFs are well below the nominal scaling law threshold due to the interaction of PDF tail distributions. To accurately constrain these tail interactions, convergence tests show that many ($\sim1e6$) samples are needed for $L$ to accurately resolve the lower tail of that distribution. Luckily, this is only needed once and thus does not have a significant impact on the computational time required for exploring Monte Carlo universes.
All $E$ distributions explicitly include the maximum possible EF instance (all $A=1, \Theta=0$) to accurately form the high end tail.
Convergence testing found that high accuracy is possible with as low as $1e4$. 
Uncertainty in the risk evaluation can be quantified by repeating the same Monte Carlo method $M$ times to report both the mean and standard deviation. Here, risks and uncertainties are reported using $M=30$.

\begin{figure}[h]
\centering{}\includegraphics[width=1\linewidth]{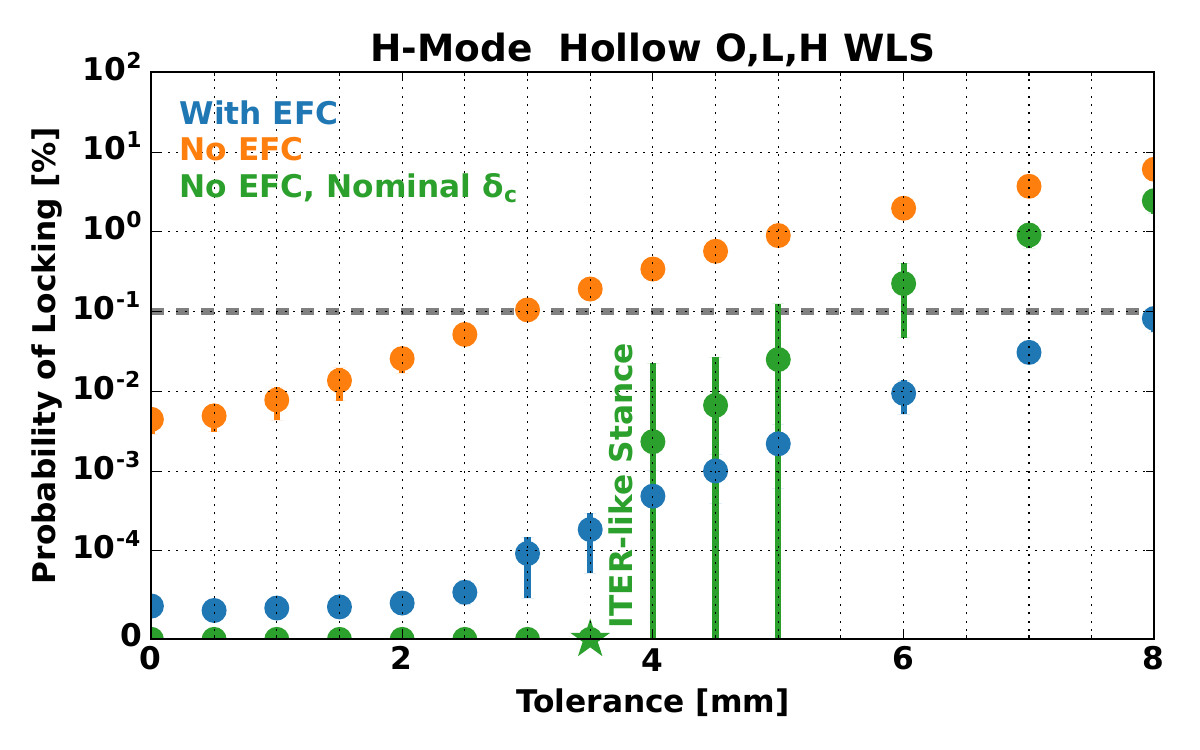}\protect\caption{The H-mode target locking risk as a function of equal shift and tilt tolerance levels for PF, DIV, and VS coils at fixed 1 mm CS tolerances. \label{fig:risk_vs_tolerance_h}}
\end{figure}

\begin{figure}[h]
\centering{}\includegraphics[width=1\linewidth]{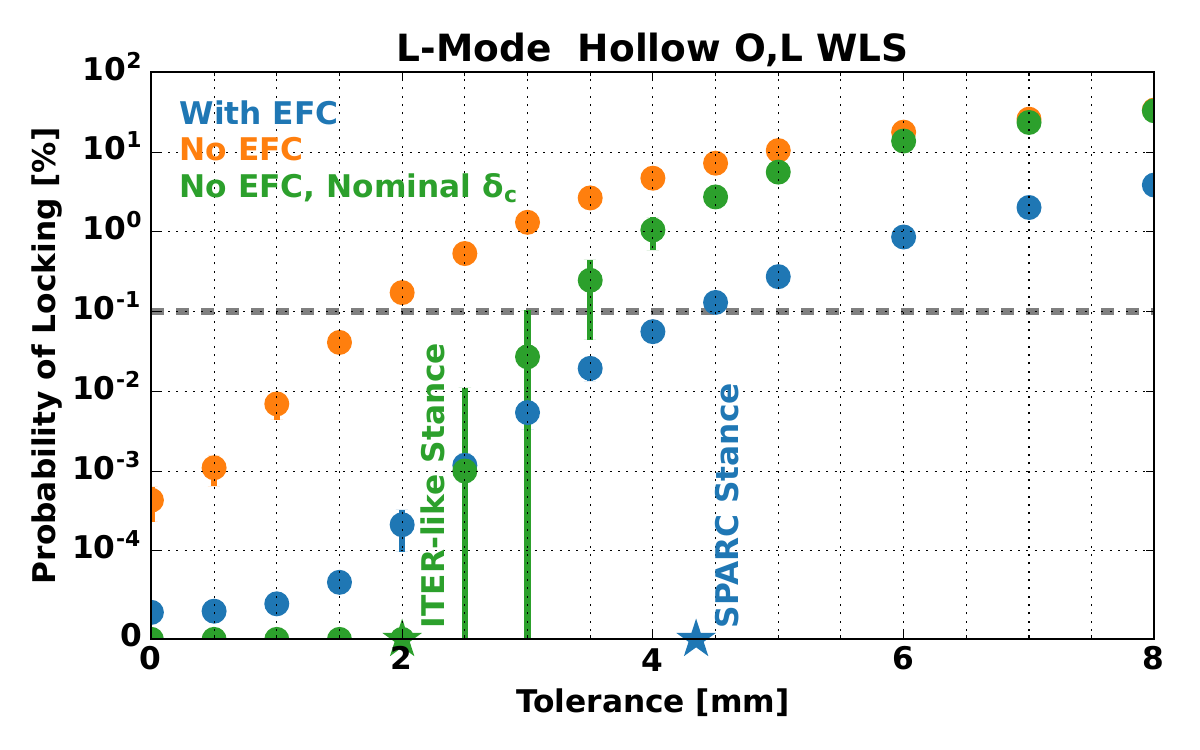}\protect\caption{The L-mode target locking risk as a function of equal shift and tilt tolerance levels for PF, DIV, and VS coils at fixed 1 mm CS tolerances. \label{fig:risk_vs_tolerance_l}}
\end{figure}

This risk quantification framework allows the efficient iterative assessment of the impact that coil design and assembly tolerance decisions have on the locking risk. For example, various tolerance schemes can be quickly mapped to risk as shown in figures \ref{fig:risk_vs_tolerance_h} and \ref{fig:risk_vs_tolerance_l}. These show the risk of locking in the target H- and L-mode scenarios respectively, for a tolerance scheme with fixed CS tolerances of 1 mm but varied tolerances on the other coils (as always, the fixed set of coherent displacement tolerances are also included). The shift and tilt tolerances are kept equal throughout these scans. 
A horizontal dashed line marks a 1-in-1000 locking risk level, which is the risk posture adopted to balance engineering feasibility with physics/operation margin on the SPARC project.

The H-mode case shows this level of risk, taking into account the EFC efficiency assumptions, is not reached until over 8 mm tolerances are allowed for assembly of the non-CS coils. 
Such a high tolerance would allow for rapid assembly and would rarely constrain coils beyond their structural tolerances (i.e. simply fitting the machine together).
This is juxtaposed with the green curve, which uses a step function threshold CDF (i.e. the nominal threshold scaling value). 
ITER uses this nominal scaling value approach and targets zero risk, meaning the worst possible assembly must be below the nominal scaling law threshold.
In this example tolerance scheme, this ITER-like approach would demand a much more strict tolerance of 3.5 mm.

The L-mode scenario has a lower locking threshold PDF as shown in \fref{fig:scaling_pdfs}, which translates to larger risk at smaller tolerances as shown in \fref{fig:risk_vs_tolerance_l}. For this scenario, the risk exceeds the 1-in-1000 threshold at 4.3 mm tolerances for the non-CS coils. 
The ITER-like scheme would demand a lower tolerance of 2.0 mm, which is similar to the PF tolerances for ITER \cite{Amoskov2015OptimizationCoils, Simon2016AssemblyTokamak, Liao2022Completion6} and low enough that it would have a significant impact on the cost and time needed for assembly.

\begin{table}[t]
  \centering
  \def\arraystretch{1.5}
  \begin{tabular}{|l|c|}
    \hline
    Coil(s) & Tol.~[mm] \\
    \hline
CS (coherent)  & 0.5 \\
VS (coherent)  & 2.0  \\
CS (individual)   & 1.0 \\
    \hline
PF (individual)   & $\leq$4.3 \\
VS (individual)   & $\leq$4.3 \\
DIV (individual)   & $\leq$4.3 \\
    \hline
  \end{tabular}
  \caption{\label{tab:tolerance_scheme}Example SPARC tolerance distribution adhering to the 1-in-1000 risk posture. The top three tolerances were prescribed and bottom three determined from  Monte Carlo scans shown in \fref{fig:risk_vs_tolerance_l}. Tolerances are for both shift and tilt individually.}
\end{table}

The complete tolerance scheme set by adhering to the 1-in-1000 risk posture for both L- and H-mode plasmas is summarized in table \ref{tab:tolerance_scheme}. 
The single tolerance values given are used for both the shift and tilt tolerances individually, meaning the maximum deviation from the nominal coil location can exceed this value.
Note, this is still not an official tolerance scheme but rather a practical example of the physics based risk assessment workflow that is used to iteratively check tolerances.  
The SPARC EF model is a living model that is iteratively being updated and used to set requirements on SPARC as new coil winding tests and design modifications are made.
Eventually, it will fold in as-built and as-assembled knowledge when these measurements are available. This will allow appropriate adjustments of the tolerances for remaining degrees of freedom in assembly.

\section{\label{sec:CoilDesign} 3D Coil Design}

\subsection{\label{sec:CoilDesign_Ovelrap} Optimizing Resonant Coupling}

Knowing the most dangerous component of the error field enables the design of EF correction coils (EFCCs) that can efficiently produce (and thus correct) this spectrum. This physics based coil design process is outlined in detail in Ref. \cite{Logan2021PhysicsTokamaks}. 

\begin{figure}[h]
\centering{}\includegraphics[width=1.0\linewidth]{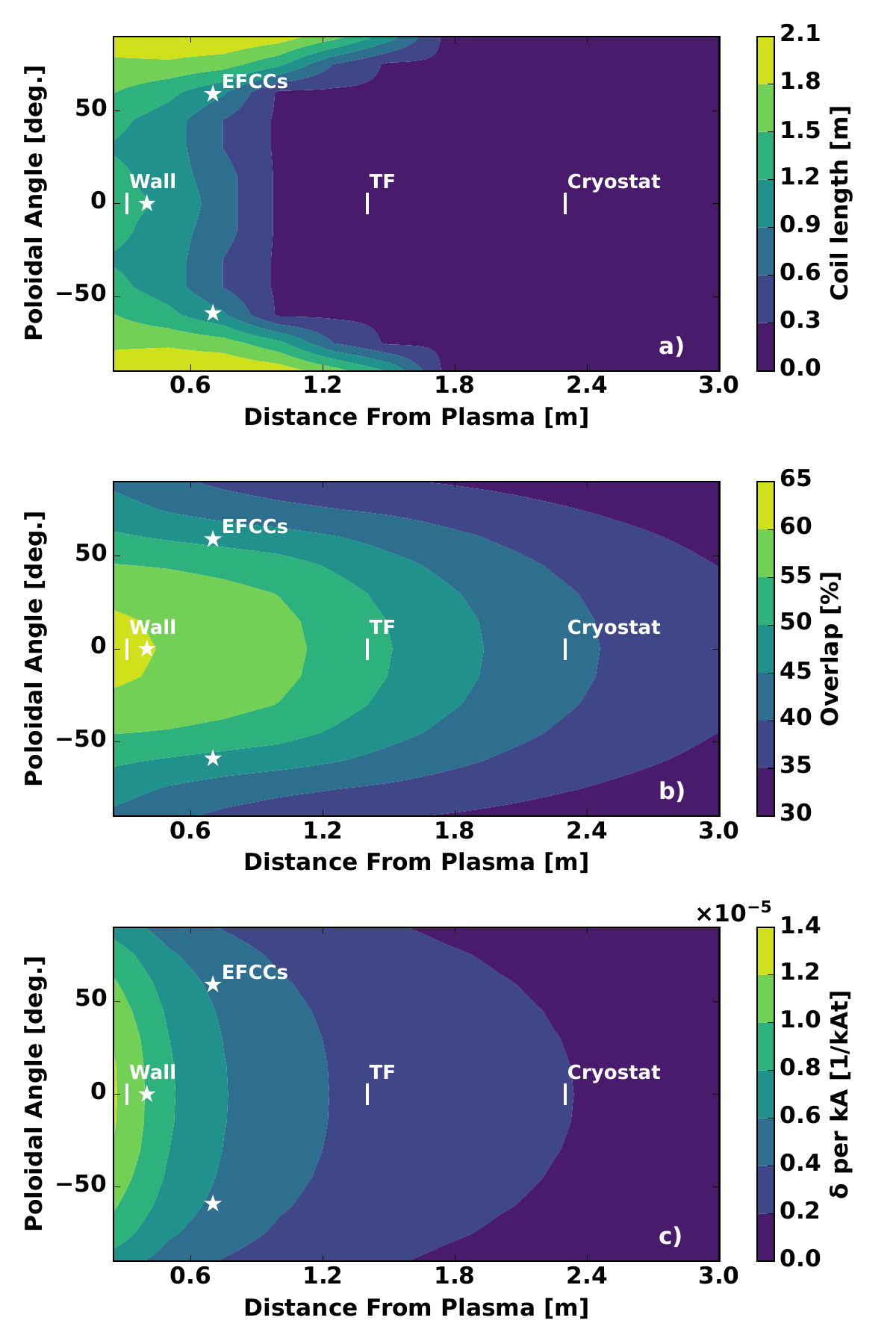}\protect\caption{Scans of potential coil arrays. (a) The optimal length of the coil poloidal cross section, (b) the corresponding overlap percentage and (b) the corresponding toroidal field normalized overlap per kA. The SPARC EFCC locations are marked by stars. Squares mark the outer surfaces of the limiter, TF coils, and cryostat for reference. \label{fig:efcc_scan}}
\end{figure}

A scan of potential coil arrays is shown in \fref{fig:efcc_scan}, providing insight into the ideal design for EFCCs. The L-mode reference equilibrium was used for this figure. An equally spaced 6 picture-frame coil toroidal array was postulated, with 6 degree gaps assumed between coils. The length of the coil poloidal cross section was scanned, as was the distance from the plasma separatrix and the poloidal angle (from the nominal plasma center) of the coil array. The poloidal tilt of the coil was kept parallel to the separatrix. 

The three subplots of \fref{fig:efcc_scan} provide an intuitive guide for optimizing coil size and placement. Subplot (a) shows the coil length corresponding to the maximum dominant mode overlap percentage ($| \tilde{\boldsymbol{\Phi}}_{x} \cdot \hat{\tilde{\boldsymbol{\Phi}}}_{c1}|/|\tilde{\boldsymbol{\Phi}}_{x}|$). This corresponds to the most purely resonant coil size for a given position. The figure shows that simple picture-frame like arrays can be reasonably expected to achieve 50-65\% resonant spectra when optimized this way, and multiple arrays are needed to achieve more purely resonant spectra. Subplot (b) shows the optimal location for resonant coils is on the midplane near the plasma, as can be intuited from \fref{fig:dominant_modes} showing the dominant mode is strongly weighted there. The optimum coil length near the plasma midplane is just over 1 meter, which can also be intuited from the midplane half wavelength in \fref{fig:dominant_modes} (see Ref. \cite{Logan2021PhysicsTokamaks} for in depth discussions of this). Close coils are further motivated by the fact that the overlap field produced per kA-turn in the coil falls off as the distance to the plasma is increased, as shown in subplot (c). 
Moving coils to the cryostat, for example, means even the optimal size coils would be majority non-resonant.
This can be compensated by adding more power supplies or turns, within engineering constraints of space, mounting stresses, and power supply capabilities. 
The loss of overlap efficiency, however, cannot be compensated for.
The smaller percent overlap, and correspondingly larger non-resonant spectral pollution, causes larger NTV torque and impacts the NTV limit of $\delta_{max}$ as demonstrated in \fref{fig:ntv_limits_lmode}.
From a physics perspective, coils closer to the plasma are more efficient.

These insights motivated low field side coils just outside of the vessel wall, mounted on the inner side of the TF coils.  Multiple coil arrays can intuitively increase the resonant coupling magnitude and efficiency by enabling poloidal spectrum tuning using the relative phase and amplitude of the $n=1$ current distributions in the arrays. Figure \ref{fig:efcc_scan} shows that multiple arrays within $\pm50^\circ$ would be preferable from an efficiency and coupling standpoint while providing the guidance that the optimal coil cross-section length off-midplane is slightly less than on-midplane.
This, combined with the vessel symmetry and simple space limitation constraints, led to the final EFCC coil design shown in \fref{fig:coil_system}.

\subsection{\label{sec:CoilDesign_FinalDesign} Designed Coil Capabilities}

\begin{figure}[h]
\centering{}\includegraphics[width=0.8\linewidth]{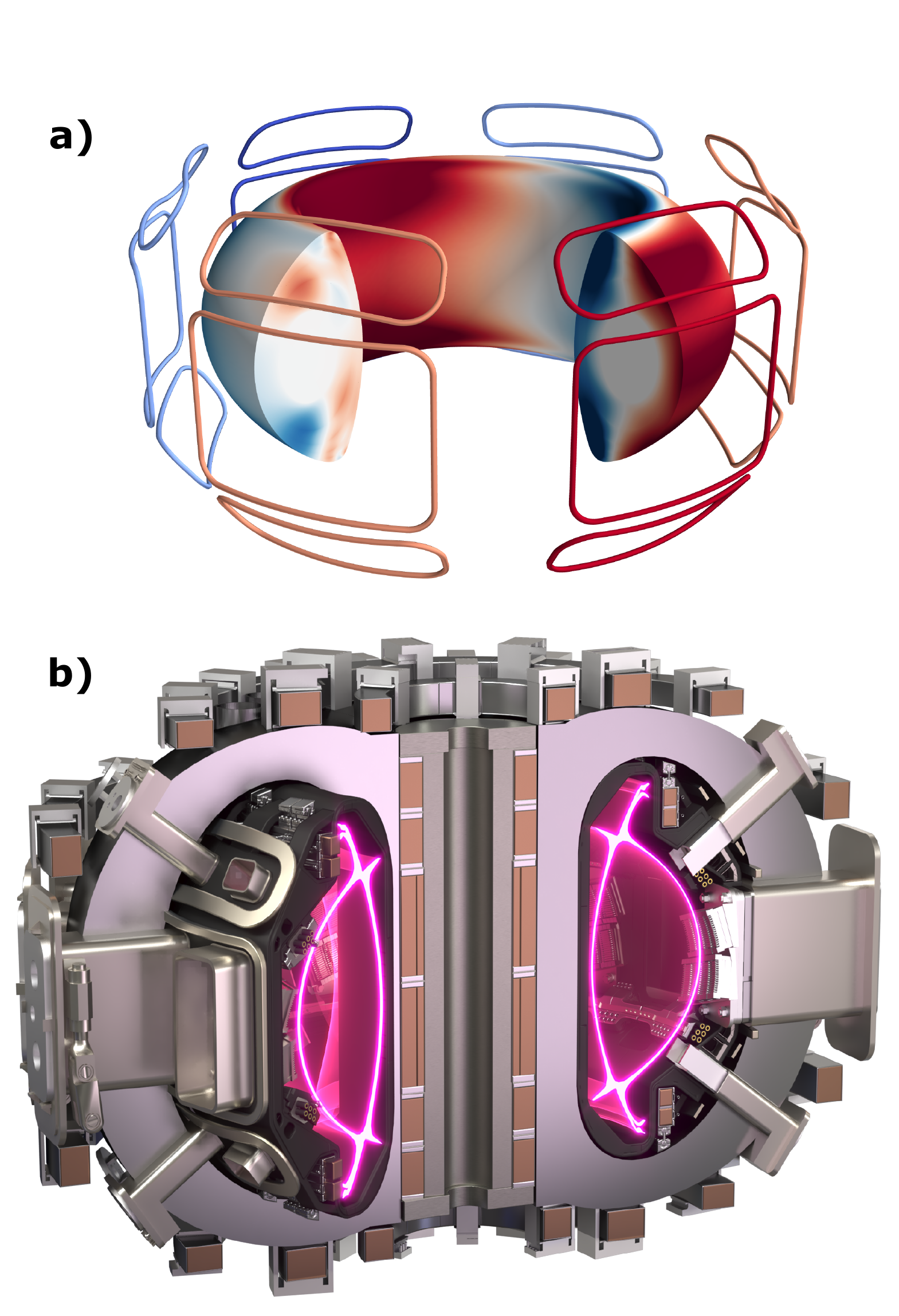}\protect\caption{The SPARC EFCC coil design: 3 rows of 6 coils each. Sub-figure (a) shows the coils in relation to the L-mode scenario plasma and the field produced from equal, in-phase currents. Sub-figure (b) shows how the EFCCs fit together with the many other hardware constraints of the device, such as port access to the vacuum  vessel.
\label{fig:EFCCs}}
\end{figure}

The SPARC EFCC system was designed in accordance with the above physics based insights and within strict engineering constraints. The EFCC system consists of 3 toroidal arrays of 6 coils each: an upper (U), midplane (M) and lower (L)  array as shown in \fref{fig:EFCCs}. Their  cross section lengths respectively match the optimal lengths in \fref{fig:efcc_scan} well given their poloidal positioning and distances from the plasma. All these values are summarized in Table \ref{tab:coil_capabilities}, which also provides the resonant overlap percent and EF per kA per turn (kAt). The coils have 100, 92, and 100 turns respectively. 
As the midplane coil is predicted to be highly effective, 42 kAt has been allocated for $n=1$ EFC with this coil. 
This corresponds to a L-mode $\delta_{EFC}=2.5e-4$, and exceeds the maximum expected from acceptably small risk tolerance schemes (as shown in \fref{fig:EF_pdfs_L-mode}, for example).
The midplane coils are designed to handle up to 150 kAt (130 kAt for U and L coils), providing capability for additional service for $n=2$ EFC or RMP ELM suppression.  

\begin{table}[t]
  \centering
  \def\arraystretch{1.5}
  \begin{tabular}{|l|c|c|c|c|c|}
    \hline
    Coil & $\theta$ & Dist. & L & $\delta$ \% & $10^{-6}\delta$/kAt \\
    \hline
U   & 59 &  0.7  &  0.7 & 44/37 & 2.3/1.9 \\
M   & 0 &  0.4  &  1.5 & 57/52 & 5.9/5.3 \\
L   & -59 &  0.7  &  0.7 & 44/38 & 2.3/1.9 \\
    \hline
  \end{tabular}
  \caption{\label{tab:coil_capabilities}SPARC EFCC array poloidal angle  [deg.], distance from the plasma [m], and cross-sectional lengths [m] as well as H-mode resonant overlap in percent and per kA-turn. Overlaps are given for L/H modes.}
\end{table}

\begin{figure}[h]
\centering{}\includegraphics[width=1\linewidth]{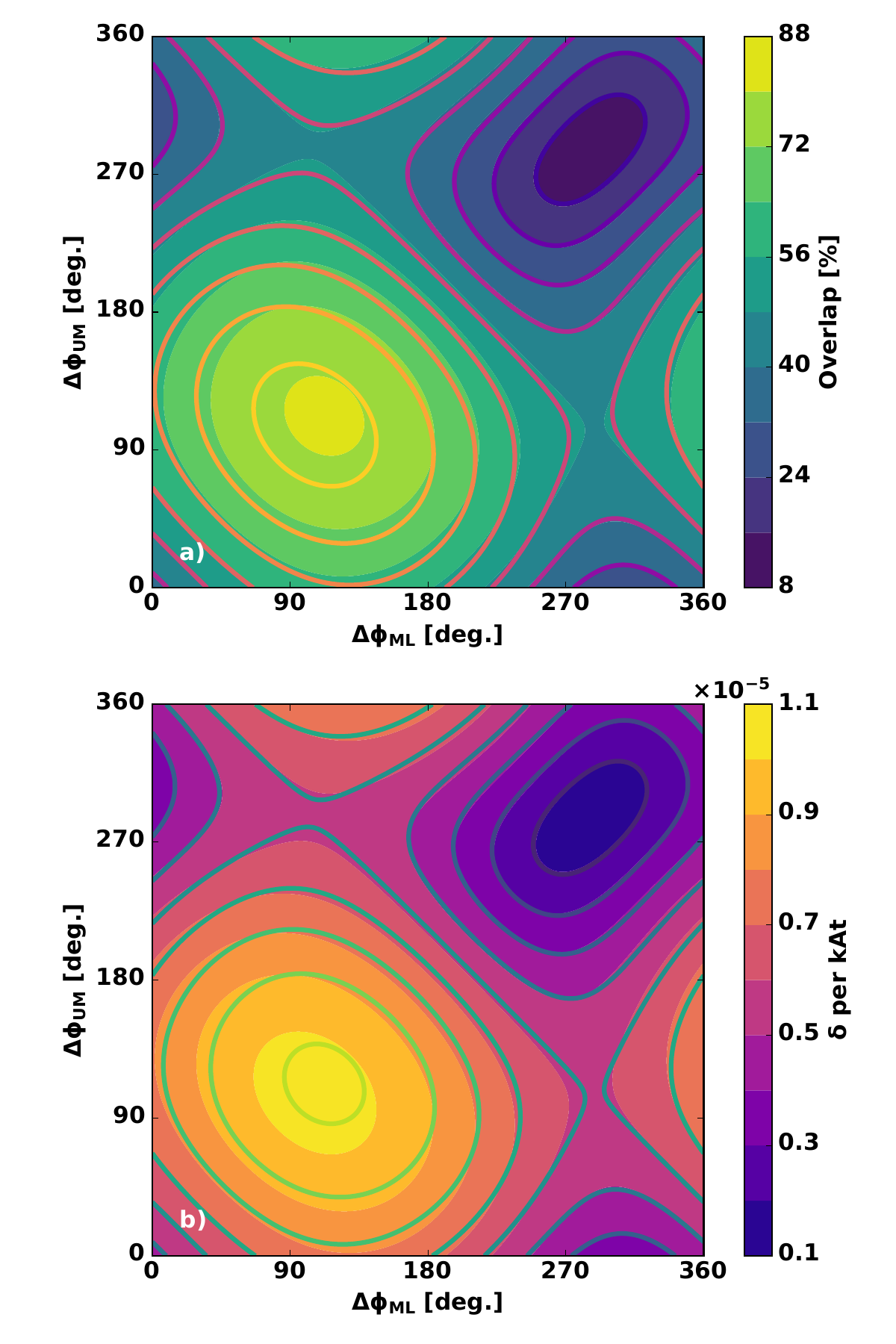}\protect\caption{Overlap percent (a) and toroidal field normalized overlap amplitude (b) for equal kAt applied in each of the U, M, and L arrays. Line contours of (a) on (b) show the efficiency and amplitude are well aligned in this space - giving a clear optimum for both near $\Delta\phi_{ML}, \Delta\phi_{UM}$ = $107^\circ, 106^\circ$.
\label{fig:phasing_scan}}
\end{figure}

If the full EFCC capabilities were to be brought to bear on the $n=1$ EFC objective, the three rows of 6 coils would enable flexible spectrum optimization to apply highly resonant $n=1$ EFC with minimal non-resonant field pollution. 
The flexibility and optimization of 3-arrays systems is often simplified by mapping the 2 phasing space ($\Delta\phi_{ML}, \Delta\phi_{UM}$) where $\Delta_{ij} = \phi_i - \phi_j$ and $\phi$ is the phase of the $n=1$ current distribution in a coil array \cite{Hu2021NonlinearBaseline, Li2017ToroidalITER}.
Such a scan is shown for the SPARC EFCCs applied to the L-mode target in \fref{fig:phasing_scan}.
Here, the amplitude in units of kAt has been fixed as equal in all 3 rows.
Even with this constraint, the optimal phasing can produce a highly core-resonant (81\% overlap) $n=1$ field.
The corresponding EFC amplitude in $\delta_{EFC} = 1.0e-5$/kAt. 
Using the full 110 kAt possible in the U and L coils, this would correspond to $\delta=1.1e-3$.
This is an order of magnitude above the nominal threshold - much larger than any acceptable tolerance scheme would allow given the conservative assumption of only $\sim 50$\% EFC efficacy and the assessment of risk based on the tail interactions of assembly and locking probabilities distributions. 
The same mapping done for the H-mode gives a similar map, with the same phasing optimum (within $3^\circ$) with only slightly reduced 71\% overlap and $\delta = 9.1e-6$/kAt. 
In conclusion, while this hyper-optimization is possible, it is expected that the MEFCCs with their 57\% resonant coupling will be more than sufficient for effective $n=1$ EF correction.
In this situation, the flexibility of these 3 rows enables the ability to use this system for  $n=2$ EFC or $n=3$ RMP ELM control should either be required.

\section{\label{sec:Summary} Summary}

This work reports the metrics used to assess the risk of locking due to intrinsic $n=1$ EFs in SPARC as well as the SPARC EFCC system's ability to correct those fields. A linear, ideal MHD core-resonance coupling is derived and singular value decomposition of the linear coupling between external fields and core resonances provides a
dominant mode spectrum that is used to identify the dangerous components of the $n=1$ EF poloidal spectrum. The overlap between a given spectrum and this dominant mode is used as the primary metric quantifying the amount of dangerous EF and an multi-device empirical scaling is used to project expectations of the locking threshold in SPARC in terms of this metric. A relatively novel risk assessment is adopted from \cite{Pharr2024ErrorITER}, the key features of which are

\begin{itemize}
    \item The large uncertainties in the empirical scaling are treated with a Monte Carlo approach to the scaling law exponents.
    \item The probability distribution of intrinsic EFs is calculated using Monte Carlo methods for a given set of prescribed tolerances (including coherent displacements of coupled coils) and nominal EF sources (feeds, magnetized seams, etc.).
    \item EFC is assumed to be only 50\% effective, a conservative choice based on past experience, empirical uncertainties, and effects expected from the non-core-resonant EFCC spectra such as including NTV.
    \item The risk of locking is calculated from the overlap of the EF PDF and locking CDF according to Eq. \eqref{eq:risk}. 
\end{itemize}

This approach is used to examine a number of reasonable tolerance schemes, showing that fixing CS coil shift and tilt tolerances to 1 mm would enable roughly 4 mm tolerances on all other coils while keeping the risk of locking below one part in one thousand. 
It is important to note that this is not a final or official tolerance scheme for SPARC. Unlike ITER's approach of setting schemes in stone \cite{Amoskov2015OptimizationCoils}, the fast timeline of SPARC will result in dynamically updated tolerances that allow reasonable trade-offs between systems while maintaining small risk.
It is also important to note, however, that the SPARC assembly plan does not plan to attempt effective EF correction through purposeful displacement of primary coils as is suggested in \cite{Zumbolo2024OptimalReduction}. 
The approach of ensuring a nominal as-designed EF far below the risk of locking risk simplifies and streamlines the requirements and communications necessary for assembly.

The SPARC EFCCs will include 3 arrays of 6 coils each with the optional ability to apply n=1,2, and 3 fields. 
The poloidal placement, distance from plasma, and size of these coil arrays matches the optimal designs well despite the many constraints on their positioning given by surrounding systems.
The $n=1$ EFC is expected to be most efficiently covered by the midplane coil, which has a 57\% match to the L-mode core resonance dominant mode spectrum and can correct most expected EFs with under 35 kAt. However, if better spectral purity or higher amplitude correction is needed the system is capable of over an 81\% spectral match and normalized EFC amplitudes exceeding $\delta = 1.1e-3$. This is far above the expected EF, but is not an over-designing of the coils because these same EFCCs have the planned secondary roles of $n=2$ EFC and/or $n=3$ RMP.

The assembly and operation of SPARC will provide critical EF measurement and EFC experience for the following ARC power plant \cite{Leuthold2025ARCBasis-MHD}. From the tokamak construction standpoint, the assembly itself will provide strategic insights into how best to utilize and balance a flexible tolerance framework while streamlining efficient assembly. From the plasma physics perspective, SPARC EF threshold measurements will be invaluable for improving the existing scalings to large, high field devices.
Quantifying the uncertainties pertaining to the efficacy of EFC in wave and alpha heated H-mode plasmas, especially those related to the NTV models that depend so heavily on the uncertain rotations in such plasmas, will also be an important result to glean from SPARC operations. This will be particularly important as future power plant scale devices like ARC will not be able to place EFC coils as close to the plasma as existing devices or SPARC.

\section{\label{sec:Acknowldgements}Acknowledgments}

This material is based on work supported by Commonwealth Fusion Systems. This work was also partially supported by the US DOE INFUSE program's 2020 project on "SPARC 3D Field Physics and Support of the Non-Axisymmetric Coil Assessment".

\printbibliography

\end{document}